\documentclass[aps,prl,twocolumn,groupedaddress]{revtex4-2}
\usepackage[utf8]{inputenc}

\usepackage{amsmath}
\usepackage{amssymb}
\usepackage{amsthm}
\usepackage{amsfonts}
\usepackage{mathtools}
\usepackage{enumerate}
\usepackage{graphicx}
\usepackage{xcolor}
\usepackage[shortlabels]{enumitem}
\usepackage{caption}
\usepackage{subcaption}
\usepackage{hyperref}
\usepackage{float}
\usepackage{array}
\usepackage{tabularx}
\usepackage{sidecap}

\usepackage{tikz}
\usetikzlibrary{arrows.meta}
\usetikzlibrary{decorations.pathreplacing}

\newtheorem*{lemma*}{Lemma}
\newtheorem*{corollary*}{Corollary}
\newtheorem*{remark*}{Remark}

\theoremstyle{definition}

\newcommand{\sym}{\mathrm{sym}}

\begin{document}

\title{Fluidization in Growth-Induced Morphogenesis}
\author{Min Wu}
\email{englier@gmail.com}
\affiliation{Department of Mathematical Sciences, Worcester Polytechnic Institute, Worcester, MA 01605, USA}
\begin{abstract}Elastic buckling has explained shape formation in growing tissues, yet the role of tissue fluidity remains elusive. We derive a minimal fluidized growth-elasticity model as a nonlinear analogue of Maxwell rheology. Analysis of a growing strip reveals a different picture of growth-induced morphogenesis: rather than emerging at a critical stress, symmetry breaking develops continuously during growth. Fluidity regulates stress evolution, the rate of shape-symmetry breaking, and flow patterns, establishing it as an active regulator of morphogenesis beyond its intuitive role in stress relaxation.\end{abstract}

\maketitle



{\it Introduction.} Fluidization is central to many biophysical processes, including embryogenesis \cite{mongera2018fluid,jain2020regionalized,tah2025minimal}, wound closure \cite{tetley2019tissue,hu2025non,jiang2026partial}, and tumor progression \cite{grosser2021cell,sauer2023changes}. At the cellular scale, fluidization arises from cell division and apoptosis in response to local stresses \cite{ranft2010fluidization}, intercellular neighbor rearrangements \cite{krajnc2018fluidization,tetley2018same,de2025epithelial}, and intracellular cytoskeletal remodeling \cite{doubrovinski2017measurement}; together, these processes continuously reorganize living elastic structures.

Growth-induced mechanical stresses and the resulting elastic buckling have been widely used to explain the emergence of complex morphologies from simple geometries \cite{amar2005growth,goriely2017mathematics,ben2025wrinkles}, such as brain convolution \cite{tallinen2014gyrification,tallinen2016growth} and gut villification \cite{ben2013anisotropic,shyer2013villification}. These descriptions generally do not account for fluidic remodeling. At sufficiently large spatial and temporal scales, tissue deformation and flow can be approximated by linear Maxwell fluids or their Stokes limit \cite{bosveld2012mechanical,streichan2018global}. 

More recently, multiple groups have incorporated viscoelasticity into models of growing biological materials \cite{garcke2022viscoelastic,Garcke2024,olaranont2025chemomechanical,zieger2026phase,wei2026continuum,slepukhin2026growth}, including our previous work with collaborators \cite{olaranont2025chemomechanical,zieger2026phase,wei2026continuum}, where phase-field simulations exhibit shape symmetry breaking arising from the combined effects of nutrient-regulated growth and atrophy, elasticity, viscoelastic remodeling, and confinement \cite{zieger2026phase}. Yet amid these multiphysical interactions, the role of fluidity, or viscoelastic remodeling, in growth-induced shape symmetry breaking remains unclear.

\begin{figure}[h]
\centering
\includegraphics[width=\columnwidth]{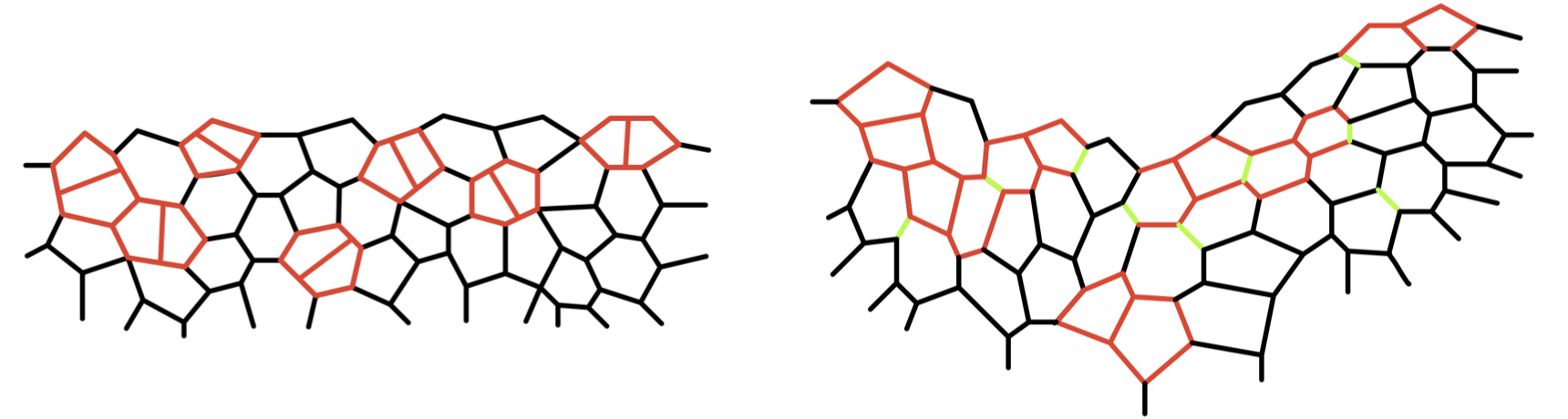}
\caption{\small Schematic of a growing tissue. Growth increases tissue mass, while cell rearrangements relax local mechanical stress. What role do they play in growth-induced shape change? Red: proliferating cells; green: new junctions.}
\label{fig0}
\end{figure}

In this manuscript, we derive a minimal, analytically tractable fluidized growth-elasticity model as a direct nonlinear analogue of Maxwell rheology, with fluidity defined as the inverse viscosity. Volumetric growth generates stress, while fluidity enables mass-conserving, isochoric rearrangements that relax growth-induced stress. Although formulated at the continuum level, this framework captures the mechanical consequence of local cell rearrangements during growth and provides a basis for understanding how cell-level fluidity regulates morphogenesis in proliferating systems and general geometric settings (Fig.~\ref{fig0}).

We show that fluidity offers different morphogenetic dynamics from the conventional elastic-buckling picture of growth-induced morphogenesis, in which growth serves as a control parameter \cite{amar2005growth,goriely2017mathematics,ben2025wrinkles,tallinen2014gyrification,tallinen2016growth,ben2013anisotropic,shyer2013villification}. Rather than emerging at a critical stress followed by post-buckling evolution, shape symmetry breaking develops continuously during growth, with fluidity regulating the baseline stress, the rate of symmetry breaking, and the resulting flow pattern. These results establish fluidity as an active, multifaceted regulator of growth-induced morphogenesis, beyond its intuitive role in relaxing stress and shifting the material response from elastic toward fluidic behavior.

{\it The model.} We consider a growing 2D strip with initial geometry
$-\infty \leq X \leq \infty$ and $0 \leq Y \leq 1$.
Driven by growth, the motion is given by
$\mathbf{x}(\mathbf{X},t)=\big(x(X,Y,t),y(X,Y,t)\big)$
with velocity field $\mathbf{v}=(x_t,y_t)$
and deformation gradient
$\mathbf{F}=\nabla_{\!X}\mathbf{x}=(\partial x_i/\partial X_j)$, where $i,j=1,2$. At any instant, the local deformation is decomposed by $\mathbf{F}=\mathbf{F}_e\mathbf{F}_g$ where $\mathbf{F}_e$ is the elastic deformation tensor and $\mathbf{F}_g$ is the growth tensor that accounts for both tissue growth and isochoric rearrangement. This multiplicative decomposition induces the additive decomposition of the velocity gradient,
$\nabla\mathbf{v}=\dot{\mathbf{F}}\mathbf{F}^{-1}
=\mathbf{\Gamma}_e+\mathbf{\Gamma}$,
where
$\mathbf{\Gamma}_e=\dot{\mathbf{F}}_e\mathbf{F}_e^{-1}$ and
$\mathbf{\Gamma}=\mathbf{F}_e(\dot{\mathbf{F}}_g\mathbf{F}_g^{-1})\mathbf{F}_e^{-1}$
are the elastic deformation-rate tensor and growth-rate tensor, respectively. Here $\nabla$ and $\nabla_X$ denote spatial and material gradients, respectively.

We model the growth-rate tensor as
$\mathbf{\Gamma}=(\gamma/2)\mathbf{I}+\mathbf{\Gamma}_D$ where the isochoric rearrangement rate is $\mathbf{\Gamma}_D=(\beta/2)\boldsymbol{\sigma}_D$,
 $\gamma$ is the areal growth rate, $\beta$ is the fluidity, and $\boldsymbol{\sigma}_D$ is the deviatoric part of the Cauchy stress $\boldsymbol{\sigma}=(\det\mathbf{F}_e)^{-1}{\partial W}/{\partial \mathbf{F}_e}\mathbf{F}_e^{T}$ with the strain energy $W(\mathbf{F}_e)$. The cell-level fluidic remodeling activities are abstracted as the isochoric rearrangement rate $\mathbf{\Gamma}_D$ in response to stress anisotropy at the continuum level, with fluidity $\beta$ controlling the rate of rearrangement per unit stress anisotropy. Together with the additive decomposition of velocity gradient, we obtain
\begin{equation}
\mathbf{D}_D=\frac{\beta}{2}\boldsymbol{\sigma}_D+\sym(\mathbf{\Gamma}_e)_D,
\label{decomposition}
\end{equation}
which is analogous to the linear Maxwell-fluid constitutive law
$\mathbf{D}_D=(\nabla\mathbf{v}+\nabla\mathbf{v}^T)/2-(\nabla\cdot\mathbf{v})/2\mathbf{I}
=\boldsymbol{\sigma}_D/(2\eta)+\dot{\boldsymbol{\sigma}}_D/(2\mu)$,
where $\eta=1/\beta$ is the viscosity and $\mu$ is the shear modulus. A thermodynamically consistent derivation is given in the Supplemental Material \cite{sm}, showing that the rearrangement term $(\beta/2)\boldsymbol{\sigma}_D$ dissipates stored elastic energy for a general strain-energy function $W(\mathbf{F}_e)$. 

For simplicity, we consider an incompressible neo-Hookean strip with strain-energy function
$W(\mathbf{F}_e)=(\mu/2)\operatorname{tr}(\mathbf{F}_e^T\mathbf{F}_e)-P(\det\mathbf{F}_e-1)$,
where $P$ is the Lagrange multiplier enforcing incompressibility $\det\mathbf{F}_e\equiv1$, equivalent to $\nabla\cdot\mathbf{v}=\gamma$ given the initial condition $\det\mathbf{F}_e(X,Y,0)=1$. Under this constitutive law, the residual
$\delta=\|\sym(\mathbf{\Gamma}_e)_D-\dot{\boldsymbol{\sigma}}_D/(2\mu)\|$
between the fluidized growth-elasticity model and the Maxwell constitutive law becomes negligible for small elastic strains. Consequently, the linearized fluidized growth-elasticity model reduces to a Maxwell fluid with a growth source. See more details in the Supplemental Material \cite{sm}.

We close the system by imposing mechanical equilibrium
$\nabla\cdot\boldsymbol{\sigma}=\mathbf{0}$ at each instant,
subject to the boundary conditions
$\mathbf{x}=\mathbf{X}$ at $Y=0$ and
$\boldsymbol{\sigma}\mathbf{n}=\mathbf{0}$ at $Y=1$,
with compatible initial conditions
$\mathbf{x}=\mathbf{X}$ and
$\mathbf{F}_g=\mathbf{I}$ everywhere at $t=0$.
Nondimensionalizing time by the growth rate,
$t'=\gamma t$,
and stress and pressure by the shear modulus,
$\boldsymbol{\sigma}'=\boldsymbol{\sigma}/\mu$ and
$P'=P/\mu$,
the system is characterized by the single dimensionless parameter
$\beta'=\beta\mu/\gamma$,
representing fluidity relative to growth and elasticity.
We study the relative fluidity $\beta'$ and henceforth drop the primes.
See the Supplemental Material \cite{sm} for details of the full model.

{\it The baseline growth-driven flow.} We first solve the system with the baseline vertical flow. From incompressibility, the symmetric solution for the motion map is given by
$x^0=X$ and $y^0=e^tY$
($y^0=e^{\gamma t}Y$ before nondimensionalization), which gives the velocity
$\mathbf{v}^0=(0,e^tY)$
and the spatially uniform strain-rate tensor
$\mathbf{D}^0=\operatorname{diag}(0,1)$
($\mathbf{D}^0=\operatorname{diag}(0,\gamma)$ before nondimensionalization). Thus, the baseline flow and strain rate are entirely determined by growth and incompressibility. 

The underlying stress dynamics is far more interesting. With fluidity $\beta>0$, the horizontal stress dynamics is given by 
\begin{equation}
\label{basestress}
\sigma_{11}^0(t)=G_1^{-1}-G_1,\text{ }
G_1(t)=
\frac{
y_1-c\,y_2e^{-\sqrt{1+\beta^2}\,t}
}{
1-ce^{-\sqrt{1+\beta^2}\,t}
},
\end{equation}
where $G_1=C_{g,11}^0$ is the horizontal (squared) growth  stretch associated with the growth Cauchy--Green tensor
$\mathbf{C}_g=\mathbf{F}_g^T\mathbf{F}_g$, $y_1=\beta^{-1}+\sqrt{1+\beta^{-2}}$, $y_2=\beta^{-1}-\sqrt{1+\beta^{-2}}$, and
$c=({1-y_1})/({1-y_2})$. 
\begin{figure}[h]
\centering
\includegraphics[width=\columnwidth]{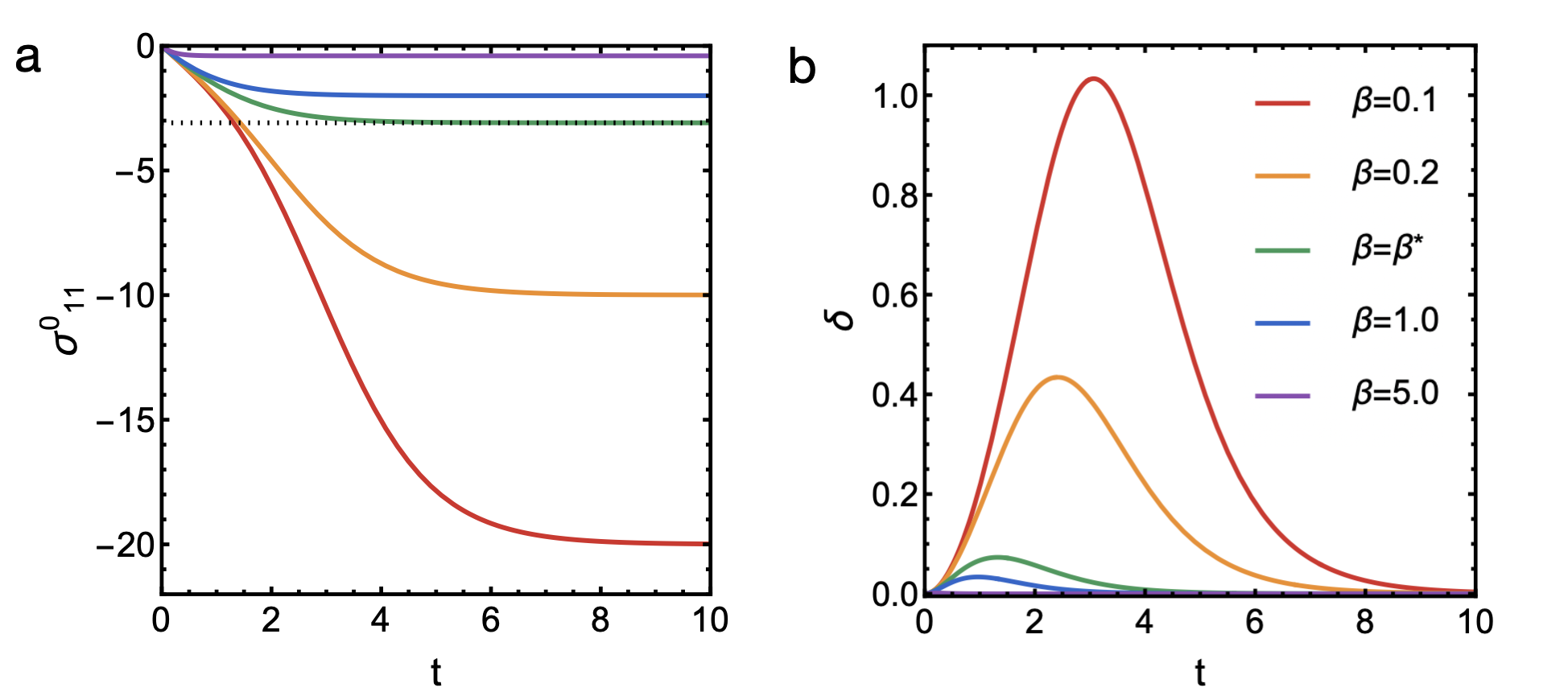}
\caption{\small Baseline horizontal stress $\sigma_{11}^{0}$ versus time for different $\beta$ (a) and deviation $\delta$  from the linear Maxwell
isochoric rheology (b). The dotted line marks the Biot threshold.}
\label{fig1}
\end{figure}

The stress has the same form as in the classical growth-elasticity model with prescribed squared growth stretch $G_1(t)$ \cite{amar2010swelling}, except that here $G_1$ is determined by the governing equations and converges to the steady state $G_{1,\mathrm{ss}}=y_1>1$. We note that Eq.(\ref{basestress}) does not apply when $\beta=0$, since in this case $G_1(t)=e^{t}$. When $\beta>0$, the horizontal stress approaches the compressive steady state
\begin{equation}
\sigma_{11,\mathrm{ss}}^{0}
=
G_{1,\mathrm{ss}}^{-1}
-
G_{1,\mathrm{ss}}
=
-\frac{2}{\beta}<0
\label{steadystress}.
\end{equation}
Fig.~\ref{fig1}a shows the dynamics of the baseline horizontal stress under different levels of fluidity. The stress monotonically approaches the steady state for each constant fluidity, with larger fluidity resulting in smaller compression at any given instant. Previous analysis showed that the symmetric solution becomes linearly unstable when the horizontal stress reaches the Biot threshold
$\sigma_{11}^{*}\sim-3.086$,
corresponding to
$G_1^{*}\sim3.38\sim1.839^2$.
This naturally motivates the definition of the critical fluidity $
\beta^{*}=-{2}/{\sigma_{11}^{*}}\sim 0.648$,
for which the steady-state stress coincides with the Biot threshold \cite{amar2010swelling}. For $\beta<\beta^{*}$, the horizontal stress reaches the Biot threshold in finite time, whereas for $\beta>\beta^{*}$ it never reaches the Biot threshold.

In conjunction with the stress dynamics, the fluidic and elastic strain
rates are
$\mathbf{\Gamma}_D^0=(\beta/2)\boldsymbol{\sigma}_D^0
=\operatorname{diag}(\beta\sigma_{11}^0/4,-\beta\sigma_{11}^0/4)$
and
$\mathbf{\Gamma}_e^0=\mathbf{D}_D^0-\mathbf{\Gamma}_D^0
=\operatorname{diag}(-(1/2+\beta\sigma_{11}^0/4),
1/2+\beta\sigma_{11}^0/4)$, respectively.
Initially, $\sigma_{11}^0=0$, so the deviatoric strain rate is entirely
elastic. As stress develops, an increasing fraction of the strain rate is
accommodated by fluidic rearrangement, reducing the rate of elastic
deformation and hence the rate of stress accumulation. At steady state,
$\sigma_{11,\mathrm{ss}}^0=-2/\beta$, such that
$\mathbf{\Gamma}_{e,\mathrm{ss}}^0=\mathbf{0}$ and
$\mathbf{\Gamma}_{D,\mathrm{ss}}^0=\mathbf{D}_D^0$, yielding a
growth-driven Stokes-flow state with
$\boldsymbol{\sigma}_D^0=(2/\beta)\mathbf{D}_D^0$ together with the
incompressibility condition. 

To quantify the departure from the linear
Maxwell-fluid limit during this transition, we evaluate
$\delta=\|\sym(\mathbf{\Gamma}_e)_D-
\dot{\boldsymbol{\sigma}}_D/(2\mu)\|$ in Fig.~\ref{fig1}b.
Starting from zero, $\delta$ first increases and then returns to zero,
capturing the transition from the linear Maxwell viscoelastic regime
of infinitesimal elastic strain and stress, through a nonlinear
viscoelastic regime, and finally to the linear fluidic regime of Stokes
flow with vanishing rates of elastic strain and stress. Overall, larger
constant fluidity accelerates this transition and leads to smaller
steady-state elastic strain and stress magnitudes. See the Supplemental Material \cite{sm} for the
baseline solution of the remaining state variables.
\begin{figure*}[t]
    \centering
    \includegraphics[width=\textwidth]{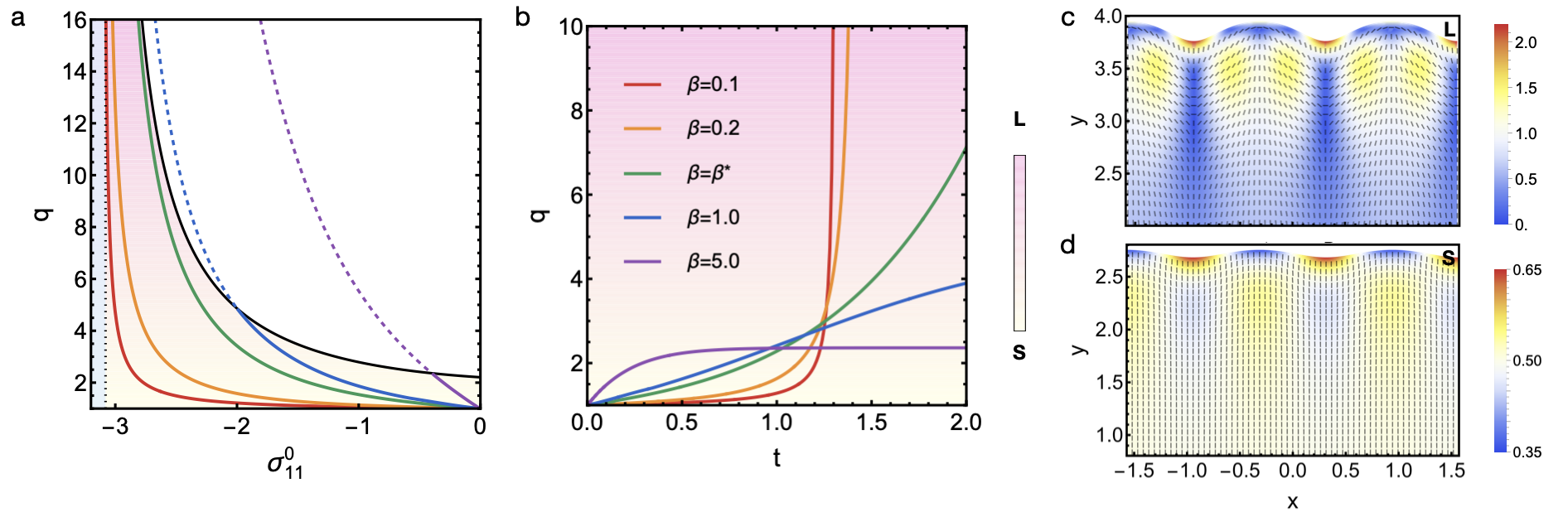}
    \caption{\small
(a) Growth rate $q$ as a function of the base-state stress $\sigma_{11}^0$ for different $\beta$. The black curve bounds the dynamically accessible regime (see text), and the dotted line denotes the Biot threshold.
(b) Evolution of $q$ in time for different $\beta$.
(c) Strain-rate pattern at $t=1.35$ for $\beta=0.2$ ($q\sim6.57$). The heat map shows the maximal extension rate ($\lambda^+$ of $\mathbf{D}_D$), and the short bars its direction. For $q\gg1$, ridges of low extension rate localize near the periodic indentations (``L''), separated by widened high-extension regions.
(d) At $t=1$ for $\beta=0.2$ ($q\sim1.63$), the strain-rate pattern is sinusoidal (``S''). The graded color from ``S'' to ``L'' in (a,b) indicates how increasing $q$ affects the strain-rate pattern.}
    \label{fig2}
\end{figure*}

{\it Linear stability analysis.}
We then study the linear stability of the baseline flow at a given time $t_0$ by introducing the displacement perturbation
$(x,y)\sim\big(X,e^t Y\big)+\epsilon e^{q(t-t_0)}\big(\cos(kX)u(Y;t_0),\sin(kX)v(Y;t_0)\big)$,
together with perturbations to pressure $P$ and growth $\mathbf{C}_g$ into the full system. See the Supplemental Material \cite{sm} for details of the resulting linear system and its solution. For the displacement, we obtain
\begin{equation}
\label{analysis}
\begin{aligned}
v(Y;t_0)=\,&
c_1e^{-\lambda_1Y}
+c_2e^{\lambda_1Y}
+c_3e^{-\lambda_2Y}
+c_4e^{\lambda_2Y},\\
u(Y;t_0)=\,&
(ke^{t_0})^{-1}{(1-q^{-1})}
v'(Y;t_0),\\
\text{with }\lambda_1=&
\frac{k e^{t_0}}{G_1(t_0)},
\lambda_2=
\frac{
k e^{t_0}
\sqrt{(q-1)+\frac{\beta}{2}\sigma_{11}^0(t_0)}
}{
\sqrt{1-q^{-1}}\,
\sqrt{(q-1)-\frac{\beta}{2}\sigma_{11}^0(t_0)}
},
\end{aligned}
\end{equation}
when $\lambda_1\neq\lambda_2$. The validity of this solution and the marginal case $\lambda_1=\lambda_2$ are discussed in the Supplemental Material \cite{sm}. 

We observe that the solution depends on the baseline horizontal compression $\sigma_{11}^0(t_0)$, or equivalently the horizontal squared growth stretch $G_1(t_0)$ through Eq.~(\ref{basestress}), the fluidity $\beta$, the rescaled wavenumber $m:=ke^{t_0}$, and the instability growth rate $q$. Substitution into the boundary conditions yields a homogeneous linear system $\mathbf{M}(c_1,c_2,c_3,c_4)^T=\mathbf{0}$, where the coefficient matrix $\mathbf{M}$ depends only on these variables. We may therefore temporarily suppress the explicit time dependence and treat $\sigma_{11}^0$, $\beta$, and $m$ as independent parameters. 

The dispersion relation $q=f_q(\sigma_{11}^0,\beta,m)$ is then determined by the singularity condition $\det\mathbf{M}=0$. The time dependence can subsequently be recovered through $\sigma_{11}^0=\sigma_{11}^0(t_0)$ and $m=ke^{t_0}$. When $\beta$ is constant in time, $\sigma_{11}^0(t_0)$ is itself determined by $\beta$ through the evolution of the base state. We first examine the dispersion relation by treating $\sigma_{11}^0$ and $\beta$ as independent parameters, before restoring their time-dependent coupling. 

In Fig.~\ref{fig2}a, we plot $q$ versus $\sigma_{11}^0$ for different values of $\beta$ in the saturated regime $m\gtrsim12$ (explained below). First, we find that for all compressive states $\sigma_{11}^0<0$ with finite fluidity $\beta>0$, the base state is linearly unstable, with $q>\gamma=1$. Intriguingly, as compression increases, the growth rate increases and blows up immediately before reaching the classical Biot threshold, indicating that fluidity promotes symmetry breaking before the elastic buckling threshold, while the Biot threshold remains effective as the limiting point with a divergent $q$, as expected for elastic buckling. Before reaching the threshold, $q$ increases with fluidity at any given level of compression.

We are able to study the combined effects of $\sigma_{11}^0$ and $\beta$ on $q$ in the saturated regime $m\gtrsim12$ because $q$ increases with $m$ and saturates for $m\gtrsim12$, as shown in Fig. S1a of the Supplemental Material \cite{sm}. Thus, for a given thickness ratio $e^{t_0}$, higher-wavenumber modes $k=m/e^{t_0}$ grow faster, while their growth rates become nearly indistinguishable beyond a threshold. This behavior echoes the elastostatic problem, where an infinite wavenumber is first selected at the Biot threshold owing to the absence of an intrinsic length scale. Such a length scale could be introduced by incorporating surface tension or gradient-dependent growth \cite{amar2010swelling}. We briefly show the effect of surface tension in Figs.~S1b,c of the Supplemental Material  \cite{sm}.


We are now positioned to consider how fluidity regulates the dynamics of $\sigma_{11}^0$ and, consequently, the instability growth rate $q$ over time (see Fig.~\ref{fig2}b). For low fluidity ($\beta<\beta^*$), the instability initially emerges slowly from the baseline growth rate ($q=\gamma=1$), accelerates as the baseline horizontal stress approaches the Biot threshold, and eventually blows up as $\sigma_{11}^0(t)\to\sigma_{11}^*$. The cases with $\beta=0.1$ and $0.2$ use the same color code as those in Fig.~\ref{fig2}a, with compression accumulating over time.

For high fluidity ($\beta>\beta^*$), see the cases with $\beta=1$ and $5$, the instability initially grows more rapidly than in the low-fluidity regime. This is consistent with their behavior in Fig.~\ref{fig2}a, where increasing fluidity increases $q$ at a given level of compression. However, as $\sigma_{11}^0$ saturates without reaching the Biot threshold $\sigma_{11}^*$, $q$ also saturates and approaches a finite plateau in time. This plateau corresponds to the intersection of each curve in Fig.~\ref{fig2}a with the boundary of the dynamically accessible region. For the critical case $\beta=\beta^*$, $q$ neither blows up nor reaches a finite plateau in finite time.

Thus, the dynamics under constant fluidity cannot access the entire parameter space in Fig.~\ref{fig2}a, where $\sigma_{11}^0$ and $\beta$ are treated as independent. Using Eq.~(\ref{steadystress}), we identify the boundary of the dynamically accessible region by imposing the steady-state relation $\beta=-2/\sigma_{11}^0$, and plot the corresponding upper bound of $q$ as a boundary in Fig.~\ref{fig2}a. Accessing the region above this boundary would require fluidity to increase from its constant baseline value at the time of perturbation, a scenario that is biophysically plausible. We further note that this boundary corresponds to the growth-driven Stokes flow for $\beta>\beta^*$.

{\it Growth-induced morphogenesis.} Although the symmetry breaking studied here is related to the elastic-buckling critical threshold, its description of growth-induced morphogenesis is fundamentally different from the conventional elastic-buckling interpretation, in which a stress threshold determines the onset of elastic buckling followed by post-buckling evolution. The fluidized theory instead suggests that growth-driven motion gradually develops undulations before reaching the elastic threshold, with the instability regulated by both stress and fluidity. Fluidity plays distinct roles over short and long times. Instantaneously, fluidity promotes the instability growth rate under accumulated compressive stress, whereas over longer times it also regulates the evolution of stress itself. For $\beta<\beta^*$, the fluidized instability accelerates as the stress approaches the Biot threshold, whereas for $\beta>\beta^*$, stress relaxation prevents the threshold from being reached and $q$ approaches a finite plateau, with the base state approaching the growth-driven Stokes flow. Thus, rather than a distinct buckling transition followed by post-buckling evolution, symmetry breaking develops continuously during growth, and fluidity can either drive an accelerating instability toward the elastic threshold or limit symmetry breaking to a finite rate.

For the flow pattern
$\mathbf{v}=(x_t,y_t)\sim (0,e^{t}Y)+\epsilon kq
\left(-\sin(kX)u(Y;t_0),\cos(kX)v(Y;t_0)\right)$,
we find a transition from sinusoidal undulations to alternating widened crests and localized indentations near the free boundary. Figure~\ref{fig2}(c,d) shows the boundary configurations and corresponding strain-rate eigenanalysis for growth with $\beta=0.2$ at an early time $t=1$ and a later time $t=1.35$, corresponding to $q\sim1.63$ and $q\sim6.57$, respectively. At the earlier, low-$q$ snapshot (Fig.~\ref{fig2}d), the alternating protruding and withdrawing boundary regions are supported by extension and contraction extending deep into the bulk. A small region immediately adjacent to the boundary exhibits the opposite trend, but is insufficient to dominate the boundary undulation. At the later, high-$q$ snapshot (Fig.~\ref{fig2}c), the withdrawing regions become strongly localized near the indentations, resembling the crease or sulcus formation previously observed in the purely elastic setting \cite{tallinen2014gyrification,tallinen2016growth,ben2025wrinkles}. 

This transition can be understood from the role of $q$ in the perturbation solution. From Eq.~(\ref{analysis}), near the free boundary $Y\sim1$, the linearized vertical motion takes the form
$v(Y;t_0)\sim c_2e^{\lambda_1Y}+c_4e^{\lambda_2Y}$,
while the lateral motion satisfies
$u(Y;t_0)\sim(1-q^{-1})
\left(G_1^{-1}c_2e^{\lambda_1Y}
+\lambda_2(ke^{t_0})^{-1}c_4e^{\lambda_2Y}\right)$.
When $q\sim1$, the lateral motion $u$ is suppressed relative to the vertical motion $v$, yielding a nearly sinusoidal flow pattern. When $q\gg1$, the lateral motion becomes significant, and its interaction with the vertical motion widens the crests and localizes the indentations. Thus, fluidity $\beta$ modulates not only the instability growth rate $q$, but also the spatial patterning of the symmetry-breaking flow through its effect on $q$. As $q$ increases away from $1$, the flow pattern gradually transitions from sinusoidal (``S'' in Fig.~\ref{fig2}) to localized (``L'' in Fig.~\ref{fig2}) indentations. 

Since this transition is directly associated with $q$, we can also compare the flow patterns across different levels of fluidity at the same time point. In Fig.~S2 of the Supplemental Material \cite{sm}, we show the flow patterns for $\beta=0.2$, $1$, and $5$ at $t=1.35$. Indeed, at the same amount of mass growth, the localization of the indentations becomes progressively weaker as fluidity increases. This suggests that low-fluidity growth may favor strongly localized, crease- or sulcus-like morphogenesis, whereas high-fluidity growth may favor smoother boundary undulations or delay the development of such localized structures. Thus, in more general and complex settings, fluidity may help determine the type of morphology that emerges during tissue folding, alongside geometric constraints and differential growth.

The problem is solved analytically up to the dispersion relation, which is evaluated numerically using Mathematica; the strain-rate eigenanalysis and visualization are also performed using Mathematica. This work is supported by NSF CAREER 2144372 and NIH R01GM157590.

\bibliography{refs}

@misc{sm,
  author = {Wu, Min},
  title = {Supplemental Material: Fluidization in Growth-Induced Morphogenesis}
}

@article{mongera2018fluid,
  title={A fluid-to-solid jamming transition underlies vertebrate body axis elongation},
  author={Mongera, Alessandro and Rowghanian, Payam and Gustafson, Hannah J and Shelton, Elijah and Kealhofer, David A and Carn, Emmet K and Serwane, Friedhelm and Lucio, Adam A and Giammona, James and Camp{\`a}s, Otger},
  journal={Nature},
  volume={561},
  number={7723},
  pages={401--405},
  year={2018},
  publisher={Nature Publishing Group UK London}
}

@article{jain2020regionalized,
  title={Regionalized tissue fluidization is required for epithelial gap closure during insect gastrulation},
  author={Jain, Akanksha and Ulman, Vladimir and Mukherjee, Arghyadip and Prakash, Mangal and Cuenca, Marina B and Pimpale, Lokesh G and M{\"u}nster, Stefan and Haase, Robert and Panfilio, Kristen A and Jug, Florian and others},
  journal={Nature communications},
  volume={11},
  number={1},
  pages={5604},
  year={2020},
  publisher={Nature Publishing Group UK London}
}

@article{tah2025minimal,
  title={A minimal vertex model explains how the amnioserosa avoids fluidization during Drosophila dorsal closure},
  author={Tah, Indrajit and Haertter, Daniel and Crawford, Janice M and Kiehart, Daniel P and Schmidt, Christoph F and Liu, Andrea J},
  journal={Proceedings of the National Academy of Sciences},
  volume={122},
  number={1},
  pages={e2322732121},
  year={2025},
  publisher={National Academy of Sciences}
}

@article{tetley2019tissue,
  title={Tissue fluidity promotes epithelial wound healing},
  author={Tetley, Robert J and Staddon, Michael F and Heller, Davide and Hoppe, Andreas and Banerjee, Shiladitya and Mao, Yanlan},
  journal={Nature physics},
  volume={15},
  number={11},
  pages={1195--1203},
  year={2019},
  publisher={Nature Publishing Group UK London}
}

@article{hu2025non,
  title={Non-canonical Wnt signaling promotes epithelial fluidization in the repairing airway},
  author={Hu, Daniel Jun-Kit and Cai, Xiaoyu Tracy and Simons, Jesse and Yun, Jina and Elstrott, Justin and Jasper, Heinrich},
  journal={Nature Communications},
  volume={16},
  number={1},
  pages={4124},
  year={2025},
  publisher={Nature Publishing Group UK London}
}

@article{jiang2026partial,
  title={Partial epithelial-to-mesenchymal transition mediates profound gap closure through growth and fluidization},
  author={Jiang, Han and Wei, Chaozhen and Wang, Pengbo and Johnson, Jaivarsini and Olaranont, Nonthakorn and Gu, Yifan and Chen, Feiyang and Xu, Jian and Wen, Qi and Wu, Min and others},
  journal={bioRxiv},
  pages={2026--07},
  year={2026},
  publisher={Cold Spring Harbor Laboratory}
}

@article{grosser2021cell,
  title={Cell and nucleus shape as an indicator of tissue fluidity in carcinoma},
  author={Grosser, Steffen and Lippoldt, J{\"u}rgen and Oswald, Linda and Merkel, Matthias and Sussman, Daniel M and Renner, Fr{\'e}d{\'e}ric and Gottheil, Pablo and Morawetz, Erik W and Fuhs, Thomas and Xie, Xiaofan and others},
  journal={Physical Review X},
  volume={11},
  number={1},
  pages={011033},
  year={2021},
  publisher={APS}
}

@article{sauer2023changes,
  title={Changes in tissue fluidity predict tumor aggressiveness in vivo},
  author={Sauer, Frank and Grosser, Steffen and Shahryari, Mehrgan and Hayn, Alexander and Guo, Jing and Braun, J{\"u}rgen and Briest, Susanne and Wolf, Benjamin and Aktas, Bahriye and Horn, Lars-Christian and others},
  journal={Advanced Science},
  volume={10},
  number={26},
  pages={2303523},
  year={2023},
  publisher={Wiley Online Library}
}

@article{tetley2018same,
  title={The same but different: cell intercalation as a driver of tissue deformation and fluidity},
  author={Tetley, Robert J and Mao, Yanlan},
  journal={Philosophical Transactions of the Royal Society B: Biological Sciences},
  volume={373},
  number={1759},
  pages={20170328},
  year={2018}
}

@article{krajnc2018fluidization,
  title={Fluidization of epithelial sheets by active cell rearrangements},
  author={Krajnc, Matej and Dasgupta, Sabyasachi and Ziherl, Primo{\v{z}} and Prost, Jacques},
  journal={Physical Review E},
  volume={98},
  number={2},
  pages={022409},
  year={2018},
  publisher={APS}
}

@article{de2025epithelial,
  title={Epithelial layer fluidization by curvature-induced unjamming},
  author={De Marzio, Margherita and Das, Amit and Fredberg, Jeffrey J and Bi, Dapeng},
  journal={Physical review letters},
  volume={134},
  number={13},
  pages={138402},
  year={2025},
  publisher={APS}
}

@article{doubrovinski2017measurement,
  title={Measurement of cortical elasticity in Drosophila melanogaster embryos using ferrofluids},
  author={Doubrovinski, Konstantin and Swan, Michael and Polyakov, Oleg and Wieschaus, Eric F},
  journal={Proceedings of the National Academy of Sciences},
  volume={114},
  number={5},
  pages={1051--1056},
  year={2017},
  publisher={National Academy of Sciences}
}

@article{streichan2018global,
  title={Global morphogenetic flow is accurately predicted by the spatial distribution of myosin motors},
  author={Streichan, Sebastian J and Lefebvre, Matthew F and Noll, Nicholas and Wieschaus, Eric F and Shraiman, Boris I},
  journal={Elife},
  volume={7},
  pages={e27454},
  year={2018},
  publisher={eLife Sciences Publications, Ltd}
}

@article{ranft2010fluidization,
  title={Fluidization of tissues by cell division and apoptosis},
  author={Ranft, Jonas and Basan, Markus and Elgeti, Jens and Joanny, Jean-Fran{\c{c}}ois and Prost, Jacques and J{\"u}licher, Frank},
  journal={Proceedings of the National Academy of Sciences},
  volume={107},
  number={49},
  pages={20863--20868},
  year={2010},
  publisher={National Academy of Sciences}
}

@article{bosveld2012mechanical,
  title={Mechanical control of morphogenesis by Fat/Dachsous/Four-jointed planar cell polarity pathway},
  author={Bosveld, Floris and Bonnet, Isabelle and Guirao, Boris and Tlili, Sham and Wang, Zhimin and Petitalot, Ambre and Marchand, Rapha{\"e}l and Bardet, Pierre-Luc and Marcq, Philippe and Graner, Fran{\c{c}}ois and others},
  journal={Science},
  volume={336},
  number={6082},
  pages={724--727},
  year={2012},
  publisher={American Association for the Advancement of Science}
}

@article{tallinen2014gyrification,
  title={Gyrification from constrained cortical expansion},
  author={Tallinen, Tuomas and Chung, Jun Young and Biggins, John S and Mahadevan, Lakshminarayanan},
  journal={Proceedings of the National Academy of Sciences},
  volume={111},
  number={35},
  pages={12667--12672},
  year={2014},
  publisher={National Academy of Sciences}
}

@article{tallinen2016growth,
  title={On the growth and form of cortical convolutions},
  author={Tallinen, Tuomas and Chung, Jun Young and Rousseau, Fran{\c{c}}ois and Girard, Nadine and Lef{\`e}vre, Julien and Mahadevan, Lakshminarayanan},
  journal={Nature Physics},
  volume={12},
  number={6},
  pages={588--593},
  year={2016},
  publisher={Nature Publishing Group UK London}
}

@article{ben2013anisotropic,
  title={Anisotropic growth shapes intestinal tissues during embryogenesis},
  author={Ben Amar, Martine and Jia, Fei},
  journal={Proceedings of the National Academy of Sciences},
  volume={110},
  number={26},
  pages={10525--10530},
  year={2013},
  publisher={National Academy of Sciences}
}

@article{shyer2013villification,
  title={Villification: how the gut gets its villi},
  author={Shyer, Amy E and Tallinen, Tuomas and Nerurkar, Nandan L and Wei, Zhiyan and Gil, Eun Seok and Kaplan, David L and Tabin, Clifford J and Mahadevan, L},
  journal={Science},
  volume={342},
  number={6155},
  pages={212--218},
  year={2013},
  publisher={American Association for the Advancement of Science}
}

@article{amar2005growth,
  title={Growth and instability in elastic tissues},
  author={Amar, Martine Ben and Goriely, Alain},
  journal={Journal of the Mechanics and Physics of Solids},
  volume={53},
  number={10},
  pages={2284--2319},
  year={2005},
  publisher={Elsevier}
}

@book{goriely2017mathematics,
  title={The mathematics and mechanics of biological growth},
  author={Goriely, Alain and others},
  volume={45},
  year={2017},
  publisher={Springer New York}
}

@article{ben2025wrinkles,
  title={Wrinkles, creases, and cusps in growing soft matter},
  author={Ben Amar, Martine},
  journal={Reviews of Modern Physics},
  volume={97},
  number={1},
  pages={015004},
  year={2025},
  publisher={APS}
}

@article{garcke2022viscoelastic,
  title={Viscoelastic Cahn--Hilliard models for tumor growth},
  author={Garcke, Harald and Kov{\'a}cs, Bal{\'a}zs and Trautwein, Dennis},
  journal={Mathematical Models and Methods in Applied Sciences},
  volume={32},
  number={13},
  pages={2673--2758},
  year={2022},
  publisher={World Scientific}
}

@article{Garcke2024,
  title = {Approximation and existence of a viscoelastic phase-field model for tumour growth in two and three dimensions},
  journal = {Discrete and Continuous Dynamical Systems - S},
  volume = {17},
  number = {1},
  pages = {221--284},
  year = {2024},
  issn = {1937-1632},
  author = {Harald Garcke and Dennis Trautwein}
}

@article{olaranont2025chemomechanical,
  title={Chemomechanical regulation of growing tissues from a thermodynamically-consistent framework and its application to tumor spheroid growth: N. Olaranont et al.},
  author={Olaranont, Nonthakorn and Wei, Chaozhen and Lowengrub, John and Wu, Min},
  journal={Journal of mathematical biology},
  volume={91},
  number={3},
  pages={31},
  year={2025},
  publisher={Springer}
}

@article{zieger2026phase,
  title={A phase-field model for viscoelastic compressible tumor growth},
  author={Zieger, Luise and Wu, Min and Wei, Chaozhen and Lowengrub, John and Aland, Sebastian},
  journal={arXiv preprint arXiv:2606.30041},
  year={2026}
}

@article{wei2026continuum,
  title={Continuum modeling of fluidic and elastic flow during growth-driven wound closure in partial-EMT cell monolayers},
  author={Wei, Chaozhen and Jiang, Han and Gu, Yifan and Olaranont, Nonthakorn and Wang, Pengbo and Wen, Qi and Sun, Yubing and Wu, Min},
  journal={arXiv preprint arXiv:2607.05820},
  year={2026}
}

@article{slepukhin2026growth,
  title={Growth-Induced Transitions in Viscoelastic Matter},
  author={Slepukhin, Valentin and Hallatschek, Oskar},
  journal={arXiv preprint arXiv:2608.15320},
  year={2026}
}

@article{amar2010swelling,
  title={Swelling instability of surface-attached gels as a model of soft tissue growth under geometric constraints},
  author={Amar, Martine Ben and Ciarletta, Pasquale},
  journal={Journal of the Mechanics and Physics of Solids},
  volume={58},
  number={7},
  pages={935--954},
  year={2010},
  publisher={Elsevier}
}

@article{rodriguez1994stress,
  title={Stress-dependent finite growth in soft elastic tissues},
  author={Rodriguez, Edward K and Hoger, Anne and McCulloch, Andrew D},
  journal={Journal of biomechanics},
  volume={27},
  number={4},
  pages={455--467},
  year={1994},
  publisher={Elsevier}
}

@misc{holzapfel2002nonlinear,
  title={Nonlinear solid mechanics: a continuum approach for engineering science},
  author={Holzapfel, Gerhard A},
  year={2002},
  publisher={Kluwer Academic Publishers Dordrecht}
}

\end{document}


\title{Supplemental Material:
Fluidization in Growth-Induced Morphogenesis}
\author{Min Wu}
\email{englier@gmail.com}
\affiliation{Department of Mathematical Sciences, Worcester Polytechnic Institute, Worcester, MA 01605, USA}

\maketitle
\tableofcontents

\section{The model}
We derive fluidity within the general growth-elasticity framework in Sec.~\ref{A}. Related formulations incorporating fluidic remodeling into growth elasticity have been developed previously \cite{olaranont2025chemomechanical,zieger2026phase,wei2026continuum}. In these formulations, the fluidity parameter does not generally have the dimensions of inverse viscosity, except in \cite{wei2026continuum} (see below for a detailed comparison). Here, to isolate and analytically investigate the role of fluidity, we adopt the simplest fluidic constitutive law in Sec.~\ref{B} and the simplest incompressible neo-Hookean elastic constitutive law in Sec.~\ref{C}, and summarize the resulting dynamical PDE system in Sec.~\ref{D}. Finally, in Sec.~\ref{E}, we establish its connection to linear Maxwell-fluid dynamics and quantify the deviation through an error function.

   \subsection{Decomposition of tensorial rates}
  \label{A}
This subsection presents the preliminaries of growth–elasticity decomposition. In particular, we review the tensorial rate decomposition that follows from the multiplicative decomposition of the deformation gradient, which provides the basis for deriving fluidity and fluidic rearrangement.

Given a flow map $\mathbf{x}(\mathbf{X},t)$ and its deformation gradient
\begin{equation}
\mathbf{F}:=\frac{\partial \mathbf{x}}{\partial \mathbf{X}},
\end{equation}
the classical growth-elasticity theory \cite{rodriguez1994stress} employs the multiplicative decomposition
\begin{equation}
\label{multidecomp}
\mathbf{F}=\mathbf{F}_e\mathbf{F}_g,
\end{equation}
where $\mathbf{F}_e$ is the elastic-deformation tensor and $\mathbf{F}_g$ is the growth tensor, which accounts for volumetric growth as well as other inelastic deformations. Differentiating the multiplicative decomposition yields
\begin{equation}
\dot{\mathbf{F}}\mathbf{F}^{-1}
=
\dot{\mathbf{F}}_e\mathbf{F}_e^{-1}
+
\mathbf{F}_e
\left(
\dot{\mathbf{F}}_g\mathbf{F}_g^{-1}
\right)
\mathbf{F}_e^{-1}.
\label{eq:Fdecomp}
\end{equation}
Given the material velocity $\mathbf{v}=\partial\mathbf{x}(\mathbf{X},t)/\partial t$, using the kinematic identity
\begin{equation}
\nabla\mathbf{v}
=
\dot{\mathbf{F}}\mathbf{F}^{-1},
\label{deformrate}
\end{equation}
and defining the elastic deformation-rate tensor
 \begin{equation}\label{eq:elasticrate}\mathbf{\Gamma}_e:=\dot{\mathbf{F}}_e\mathbf{F}_e^{-1}\end{equation} 
and the growth-rate tensor
\begin{equation}
\mathbf{\Gamma}
:=
\mathbf{F}_e
\left(
\dot{\mathbf{F}}_g\mathbf{F}_g^{-1}
\right)
\mathbf{F}_e^{-1},
\label{eq:GammaDef}
\end{equation}
we obtain the additive decomposition of the velocity gradient
\begin{equation}
\mathbf{\Gamma}_e+\mathbf{\Gamma}
=
\nabla\mathbf{v},
\end{equation}
where the gradient operator $\nabla:=\nabla_{\mathbf{x}}$ is defined in the current frame. The decomposition is equivalent to the evolution of the elastic-deformation tensor
\begin{eqnarray}
\dot{\mathbf{F}}_e\mathbf{F}_e^{-1}
&=&\nabla\mathbf{v}-\mathbf{\Gamma}.
\label{eq:FeEvolution}
\end{eqnarray}
Further, we decompose the growth-rate tensor into the isotropic volumetric-growth part and the isochoric rearrangement part: \begin{equation}
\label{ratedecompose}
\mathbf{\Gamma}
=\frac{\gamma}{d}\mathbf{I}+
\mathbf{\Gamma}_D,
\end{equation}
where the volumetric growth rate $\gamma:=\tr{(\mathbf{\Gamma})}$, $d$ is the spatial dimension, and the isochoric part $\mathbf{\Gamma}_D$ is traceless. Thus, we can write the evolution of the growth tensor

\begin{equation}\dot{\mathbf{F}}_g\mathbf{F}_g^{-1}=
\frac{\gamma}{d}\mathbf{I}+\mathbf{F}_e^{-1}
\mathbf{\Gamma}_D
\mathbf{F}_e.
\label{eq:FgEvolution}
\end{equation}
We will model the isochoric part $\mathbf{\Gamma}_D$ in response to stress to derive fluidity in Sec.~\ref{B}.

To emphasize that the isochoric rate $\mathbf{\Gamma}_D$ should not contribute to volume growth, we review the volume ratios with respect to the initial state as
\begin{equation}
J=\det(\mathbf{F}), \qquad
J_e=\det(\mathbf{F}_e), \qquad
J_g=\det(\mathbf{F}_g),
\end{equation}
where the total volume ratio $J=J_eJ_g$ is the product of the volume ratio due to elastic deformation, $J_e$ and the volume ratio due to growth, $J_g$. Using Jacobi's formula and Eqs.~(\ref{deformrate}), (\ref{eq:FgEvolution}), and (\ref{eq:FeEvolution}), we obtain 
\begin{equation}
\dot{J}
=J \tr\left(\dot{\mathbf{F}}\mathbf{F}^{-1}\right)=J\,\nabla\cdot\mathbf{v},
\end{equation}
\begin{equation}
\label{eq:dotJg}
\dot{J}_g
=J_g\tr\left(\dot{\mathbf{F}}_g\mathbf{F}_g^{-1}\right)=
J_g\,\gamma,
\end{equation}
and\begin{equation}\label{eq:dotJe}
\dot{J}_e
=J_e\tr\left(\dot{\mathbf{F}}_e\mathbf{F}_e^{-1}\right)=
J_e\left(\nabla\cdot\mathbf{v}-\gamma\right).
\end{equation}
From these equations, we see clearly that the volume-ratio dynamics are independent of the isochoric rearrangement rate $\mathbf{\Gamma}_D$.

\subsection{Fluidity as the stress sensitivity of dissipative rearrangement}
\label{B}
In this subsection, we derive fluidic remodeling, or rearrangement, by modeling $\mathbf{\Gamma}_D$ in response to the Cauchy stress. We observe that the sensitivity parameter $\beta$, introduced below, has has the dimension of inverse viscosity, motivating the term \emph{fluidity}.

For an arbitrary growing region $\Omega_t$ with initial pre-growth region $\Omega_0$, where $\Omega_t=\mathbf{x}(\Omega_0)$, we define the elastic energy 
\begin{equation}
E=\int_{\Omega_0}W(\mathbf{F}_e)J_g d\mathbf{X}=\int_{\Omega_t}W(\mathbf{F}_e)J_e^{-1} d\mathbf{x}
\end{equation}
where $W(\mathbf{F}_e)$ is the elastic energy density function per unit volume in the virtual stress-free grown state. By Reynolds transport Theorem
\begin{eqnarray}
\dot{E}&=&\int_{\Omega_t}\dot{\overline{W(\mathbf{F}_e)J_e^{-1}}}+ W(\mathbf{F}_e)J_e^{-1} \nabla\cdot\mathbf{v} d\mathbf{x}\\
\nonumber &=&\int_{\Omega_t}J_e^{-1}\frac{\partial W}{\partial \mathbf{F}_e}:\dot{\mathbf{F}}_e+ W(\mathbf{F}_e)J_e^{-1} (\nabla\cdot\mathbf{v} -\dot{J_e}J_e^{-1})d\mathbf{x}\\
\nonumber &=&\int_{\Omega_t}J_e^{-1}\frac{\partial W}{\partial \mathbf{F}_e}\mathbf{F}_e^T:(\nabla\mathbf{v}-\mathbf{\Gamma})+ W(\mathbf{F}_e)J_e^{-1} \gamma d\mathbf{x}
\end{eqnarray}where the third equality uses Eq.~(\ref{eq:FeEvolution}) and Eq.~(\ref{eq:dotJe}), and $\dot{\overline\Box}:=D\Box/Dt$. Assuming that the stress is determined by the elastic response, the Cauchy stress is derived from the elastic energy
\begin{equation}\boldsymbol\sigma = J_e^{-1}\frac{\partial W}{\partial \mathbf{F}_e}\mathbf{F}_e^T
\end{equation}
and the corresponding first Piola-Kirchhoff stress is 
\begin{equation}\boldsymbol\Pi = J_g\frac{\partial W}{\partial \mathbf{F}_e}\mathbf{F}_g^{-T}
\end{equation}
which follows from the Piola transformation $\boldsymbol\Pi =J\boldsymbol\sigma\mathbf{F}^{-T}$. Thus, \begin{eqnarray}
\dot{E} &=&\int_{\Omega_t}\boldsymbol\sigma:\nabla\mathbf{v}-\boldsymbol\sigma:\mathbf{\Gamma}+ W(\mathbf{F}_e)J_e^{-1} \gamma d\mathbf{x},
\end{eqnarray} where the first term is the stress power due to the observable deformation and the last term is the change in elastic energy associated with volumetric growth. The middle term, $-\boldsymbol\sigma:\mathbf{\Gamma}$ is the stress power associated with inelastic growth and rearrangement. If $-\boldsymbol\sigma:\mathbf{\Gamma}>0$, elastic energy is stored due to growth; if $-\boldsymbol\sigma:\mathbf{\Gamma}<0$, elastic energy is dissipated due to growth.

Using Eq.~(\ref{ratedecompose}), we decompose
\begin{equation}
-\boldsymbol{\sigma}:\mathbf{\Gamma}
=
-\boldsymbol{\sigma}_D:\mathbf{\Gamma}_D
-\frac{\tr(\boldsymbol{\sigma})}{d}\mathbf{I}:\frac{\gamma}{d}\mathbf{I}
=
-\boldsymbol{\sigma}_D:\mathbf{\Gamma}_D
-\gamma\frac{\tr(\boldsymbol{\sigma})}{d}.
\end{equation}
Thus, $-\boldsymbol{\sigma}_D:\mathbf{\Gamma}_D$ represents the stress power associated with isochoric rearrangement, whereas $-\gamma\tr(\boldsymbol{\sigma})/d$ represents the stress power associated with volumetric growth. In principle, both $\gamma$ and $\mathbf{\Gamma}_D$ may depend on the mechanical state. Here, we treat $\gamma$ as a prescribed global growth rate and focus on stress-driven isochoric rearrangement. When the isochoric rearrangement dissipates elastic energy 
$-\boldsymbol{\sigma}_D:\mathbf{\Gamma}_D\le0$,
in a rate-dependent manner, it motivates a family of constitutive laws that satisfy this inequality. For simplicity, we choose
\begin{equation}\label{eq:fluidization}
\mathbf{\Gamma}_D=\frac{\beta}{2}\boldsymbol{\sigma}_D,
\end{equation}
where $\beta\ge0$ is defined as the fluidity with the physical dimension of inverse viscosity. In Sec.~\ref{E}, we show that when $\beta>0$, $1/\beta$ corresponds to the viscosity in the Maxwell-fluid limit. 
This constitutive law ensures dissipation through  \begin{equation}-\boldsymbol\sigma_D:\mathbf{\Gamma}_D=-\frac{\beta}{2} \boldsymbol\sigma_D:\boldsymbol\sigma_D\leq 0.\end{equation} A more sophisticated dissipative constitutive law incorporating cell-monolayer polarity was considered in \cite{wei2026continuum}. In related formulations \cite{olaranont2025chemomechanical,zieger2026phase}, fluidity was instead introduced as a rate $\beta_0$, with $\mathbf{\Gamma}_D=\beta_0(\mathbf{F}_e\mathbf{F}_e^T)_D$.

Here, the evolution Eqs.(\ref{eq:FeEvolution}) and (\ref{eq:FgEvolution}) become
\begin{equation}
\dot{\mathbf{F}}_e\mathbf{F}_e^{-1}
=\nabla\mathbf{v}-\left(\frac{\gamma}{d} \mathbf{I}+\frac{\beta}{2}\boldsymbol\sigma_D\right)
\label{eq:FeEvolution1}
\end{equation}
and
\begin{equation}
\dot{\mathbf{F}}_g\mathbf{F}_g^{-1}
=
\frac{\gamma}{d}\mathbf{I}
+
\frac{\beta}{2} \mathbf{F}_e^{-1}
\boldsymbol\sigma_D
\mathbf{F}_e
\label{eq:FgEvolution1}
\end{equation}
respectively. 
\subsection{Soft tissues as an incompressible neo-Hookean material}
\label{C} 
In this subsection, we close the system using the simplest neo-Hookean material \cite{holzapfel2002nonlinear}. Most previous analytical studies impose incompressibility, since the resulting symmetric baseline state can often be obtained in a simple closed form. Since our primary goal is to provide analytical insight into the role of fluidization in growth-induced symmetry breaking, we choose an incompressible neo-Hookean material, whose elastic energy density is given by
\begin{equation}
W(\mathbf{F}_e) = \frac{\mu}{2}\left(\mathbf{F}_e:\mathbf{F}_e\right)-P\left(J_e-1\right)
\end{equation}
with the corresponding Cauchy stress 
\begin{equation}
\label{eq:cauchystress}
\boldsymbol\sigma =\mu \mathbf{F}_e\mathbf{F}_e^T-P \mathbf{I}
\end{equation}
and the first Piola-Kirchhoff stress

\begin{equation}\boldsymbol\Pi =  J_g\left(\mu\mathbf{F}_e-P \mathbf{F}_e^{-T}\right)\mathbf{F}_g^{-T}\label{eq:pk1stress}
\end{equation} 
where $\mu$ is the shear modulus and the Lagrange multiplier $P$ enforces the incompressible constraint \begin{equation}J_e\equiv 1.
 \end{equation} 
This condition is equivalent to prescribing the initial condition $J_e(\mathbf{X},t=0)=1$ together with $\dot{J}_e=0$. Using Eq.~(\ref{eq:dotJe}), we obtain the equivalent condition
\begin{equation}
\left\{
\begin{aligned}
J_e(\mathbf X,0)&=1,\\
\nabla\cdot\mathbf v&=\tr(\dot{\mathbf{F}}\mathbf{F}^{-1})=\gamma
\end{aligned}
\right.
\quad \Longleftrightarrow \quad
J_e(\mathbf X,t)\equiv 1
\end{equation}
which ensures that the elastic deformation remains incompressible in time.

\subsection{Model summary and nondimensionalization}
\label{D}

Before we list the full system, we introduce the right Cauchy-Green tensor of growth $\mathbf{C}_g:=\mathbf{F}_g^T\mathbf{F}_g$ from which (\ref{eq:pk1stress}) can be rewritten as

\begin{equation}\label{stresspk}\boldsymbol\Pi = J_g\left(\mu\mathbf{F}\mathbf{C}_g^{-1}-P \mathbf{F}^{-T}\right),
\quad\text{with } J_g =(\det{\mathbf{C}_g})^{1/2}.
\end{equation}

Using Eqs. (\ref{multidecomp}), (\ref{eq:FgEvolution1}), and (\ref{eq:cauchystress}), the evolution of $\mathbf{C}_g$ is given by

\begin{equation}
\dot{\mathbf{C}}_g\mathbf{C}_g^{-1}
=
\frac{2\gamma}{d}\mathbf{I}
+
\beta \mu \left(\mathbf{C}\mathbf{C}_g^{-1}-\frac{\tr\left(\mathbf{C}\mathbf{C}_g^{-1}\right)}{d}\mathbf{I}
\right) \label{eq:CgEvolution1}
\end{equation}
where $\mathbf{C}:=\mathbf{F}^T\mathbf{F}$ is the right Cauchy-Green tensor of deformation. By right-multiplying $\mathbf{C}_g$ on both sides of the equation, one can see the symmetry of $\mathbf{C}_g$ is maintained along its trajectory given a proper symmetric initial condition. Since the stresses depend on $\mathbf{C}_g$ directly, solving Eq.(\ref{eq:CgEvolution1}) is sufficient. Analytically, this is more efficient than solving the nonsymmetric Eq.(\ref{eq:FgEvolution1}).

 In addition, similar as Eq.(\ref{eq:FgEvolution1}) implying Eq.(\ref{eq:dotJg}), the evolution of ${\mathbf{C}}_g$ implies \begin{equation}\tr(\dot{\mathbf{C}}_g\mathbf{C}_g^{-1})=2\gamma \quad \Longleftrightarrow\quad \dot{\overline{\det\mathbf{C}}}_g = 2\gamma \det\mathbf{C}_g,\label{implied_non-dimensional}\end{equation}
where the isochoric rearrangement does not affect the volume ratio.

Next, we nondimensionalize time by the growth time scale $\gamma^{-1}$,
$t'=\gamma t$, and scale the stress and pressure by the shear modulus,
$\boldsymbol{\sigma}'=\boldsymbol{\sigma}/\mu$ and $P'=P/\mu$.
The fluidity is then rescaled as
$\beta'=\beta\mu/\gamma$.
Primes are omitted hereafter. The full nondimensionalized system in the reference configuration is
given by
\begin{empheq}[left=\empheqlbrace]{align}
\label{thesystem1}
&\nabla_0\cdot\mathbf{\Pi}=\mathbf{0},
\quad
\text{with }\quad\mathbf{\Pi}
=
J_g\left(
\mathbf{F}\mathbf{C}_g^{-1}
-P\,\mathbf{F}^{-T}
\right),
\quad J_g =(\det{\mathbf{C}_g})^{1/2}\text{ and }\quad
\mathbf{F}=\nabla_0\mathbf{x},
\\
&\nabla\cdot\mathbf v=\tr\!\left(\dot{\mathbf{F}}\mathbf{F}^{-1}\right)=1,\label{thesystem2}
\\
&\dot{\mathbf{C}}_g\mathbf{C}_g^{-1}
=
\frac{2}{d}\mathbf{I}
+
\beta \left(\mathbf{C}\mathbf{C}_g^{-1}-\frac{\tr\left(\mathbf{C}\mathbf{C}_g^{-1}\right)}{d}\mathbf{I}
\right) ,\qquad
\mathbf{C}=\mathbf{F}^T\mathbf{F}
\label{thesystem3}
\end{empheq}
where the motion map $\mathbf{x}(\mathbf{X},t)$,  the pressure $P(\mathbf{X},t)$, and the right Cauchy-Green growth tensor $\mathbf{C}_g(\mathbf{X},t)$ are the unknown fields. All the other state variables such as the Cauchy stress $\boldsymbol\sigma$ can be computed from these state variables. $\nabla_0$ is the gradient operator in the reference coordinate. The right-hand side of Eq.(\ref{thesystem2}) is unity because time has been scaled by $\gamma$. 
The initial conditions are
\begin{equation}
\left\{
\begin{aligned}
&\mathbf{x}(\mathbf{X},0)=\mathbf{X},
\\
&\mathbf{C}_g(\mathbf{X},0)=\mathbf{I},
\end{aligned}
\right.
\end{equation}
and the boundary conditions are
\begin{equation}
\left\{
\begin{aligned}
&\mathbf{x}(\mathbf{X},t)=\mathbf{X}
\qquad \text{on } \partial\Omega_0^d,
\\
&\mathbf{\Pi}\mathbf{N}=\mathbf{0}
\qquad \text{on } \partial\Omega_0^n,
\label{fullbc}
\end{aligned}
\right.
\end{equation}
where $\partial\Omega_0^d$ and $\partial\Omega_0^n$ denote the fixed and
free parts of the boundary, respectively, such that
$\partial\Omega_0=\partial\Omega_0^d\cup\partial\Omega_0^n$, $\partial\Omega_0^d\cap\partial\Omega_0^n=\varnothing$, and
$\mathbf{N}$ is the outward unit normal on $\partial\Omega_0^n$. 

%
\subsection{Comparison with linear incompressible Maxwell fluids}
\label{E}
%

This subsection demonstrates that coupling isochoric rearrangement to the traceless stress through Eq.~(\ref{eq:fluidization}) fluidizes the growth-elasticity framework in the Maxwell sense. For comparison, we will use the dimensional system. 

First of all, the deviatoric part of the constitutive law is given by 
\begin{equation}
\label{eq:nonlinearM}
\mathbf{D}_D = \frac{\boldsymbol\sigma_D}{2\eta} +\frac{\dot{\boldsymbol\sigma}_D}{2 \mu'},
\end{equation}
where $\mathbf{D}_D:=\sym(\nabla \mathbf{v})_D=(\nabla \mathbf{v}+\nabla \mathbf{v}^T)/2-({\nabla\cdot\mathbf{v}})/{d}\mathbf{I}$, $\eta$ is the viscosity, and $\mu'$ is the elastic modulus. The normal stress acts as a pressure $\tilde{P}$ associated with the constraint $\nabla\cdot\mathbf{v}=\gamma$.

Using the Eqs.~(\ref{eq:elasticrate}), (\ref{eq:FeEvolution}), and (\ref{eq:fluidization}), we have 
\begin{equation}
\label{eq:additive0}
\nabla \mathbf{v} = \frac{\gamma}{d}\mathbf{I}+\frac{\beta}{2} \boldsymbol\sigma_D +\mathbf{\Gamma}_e,
\end{equation}
which leads to the additive decomposition of the traceless part of the strain rate tensor
\begin{equation}
\label{eq:additive}
\mathbf{D}_D = \frac{\beta}{2} \boldsymbol\sigma_D +\sym(\mathbf{\Gamma}_e)_D. 
\end{equation}

When $\sym(\mathbf{\Gamma}_e)_D = \dot{\boldsymbol\sigma}_D/(2\mu)$, with viscosity $\eta=1/\beta$ and elastic modulus $\mu'=\mu$,  Eq.~(\ref{eq:additive}) is formally equivalent to Eq.~(\ref{eq:nonlinearM}).
We therefore define the error
\begin{equation}
\delta
=
\left\|
\sym(\mathbf{\Gamma}_e)_D
-
\frac{\dot{\boldsymbol\sigma}_D}{2\mu}
\right\|
\end{equation}
to quantify the deviation of the current model from the Maxwell model at any instant.

Introducing the left Cauchy--Green tensor of elastic deformation $\mathbf B_e:=\mathbf F_e\mathbf F_e^T$ and using Eq.~(\ref{eq:cauchystress}), we obtain
 \begin{equation}
\label{eq:stressD}
\boldsymbol\sigma_D=\mu \left(\mathbf{B}_e-\frac{\tr\left(\mathbf{B}_e\right)}{d}\mathbf{I}\right)=\mu (\mathbf{B}_e)_D.
\end{equation}Together with Eq.~(\ref{eq:elasticrate}), we have
\begin{equation}
\frac{\dot{\boldsymbol{\sigma}}_D}{2\mu}
=
\frac{1}{2}(\dot{\mathbf{B}}_e)_D
=
\sym(\mathbf{\Gamma}_e\mathbf{B}_e)_D.
\end{equation}
Therefore,
\begin{eqnarray}
\label{eq:delta}
\delta
&=&
\left\|
\sym(\mathbf{\Gamma}_e)_D
-
\sym(\mathbf{\Gamma}_e\mathbf{B}_e)_D
\right\|
\nonumber\\
&=&
\left\|
\sym\left(
\mathbf{\Gamma}_e(\mathbf{I}-\mathbf{B}_e)
\right)_D
\right\|.
\end{eqnarray}

When the elastic strain is small such that
\[
\mathbf F_e=\mathbf I+\epsilon\mathbf f_e,
\qquad |\epsilon|\ll1,
\]
we have
\[
\mathbf B_e-\mathbf I
=
\epsilon(\mathbf f_e+\mathbf f_e^T)
+o(\epsilon),
\qquad
\mathbf\Gamma_e
=
\epsilon\dot{\mathbf f}_e
+o(\epsilon),
\]
and therefore
\[
\delta=O(\epsilon^2).
\]
In this case, the discrepancy between Eq.~(\ref{eq:additive}) and Eq.~(\ref{eq:nonlinearM}) is second order in the elastic strain. Thus, the linearized model recovers the deviatoric stress dynamics of the linear Maxwell fluid.

For completeness, we list the full linearized model. Since the normal stress dynamics is enslaved by the pressure term due to incompressibility, the full system is given by
\begin{equation}
\nabla\cdot\boldsymbol\sigma =\nabla\cdot\boldsymbol\sigma_D-\nabla \tilde P =0, \quad  \nabla\cdot\mathbf{v}=\gamma, \quad \dot{\boldsymbol\sigma}_D= \mu(2\mathbf{D}_D-{\beta} \boldsymbol\sigma_D)={\mu}\left(\nabla \mathbf{v}+\nabla \mathbf{v}^T-\frac{2}{d}\left(\nabla\cdot\mathbf{v}\right)\mathbf{I}\right)-{\beta\mu} \boldsymbol\sigma_D,
\end{equation}
with the initial condition
\begin{equation}
\boldsymbol\sigma_D =\mathbf{0} \quad\text{at } t=0
\end{equation}
and the boundary condition
\begin{equation}
\mathbf{v}(\mathbf{x},t)=\mathbf{0}
\quad \text{on } \partial\Omega_t^d, \quad\text{and}\quad \boldsymbol{\sigma}\mathbf{n}=\mathbf{0}
\quad \text{on } \partial\Omega_t^n
\end{equation}
where $\partial\Omega_t^d$ and $\partial\Omega_t^n$ denote the fixed and
free parts of the boundary, respectively, such that
$\partial\Omega_t=\partial\Omega_t^d\cup\partial\Omega_t^n$, $\partial\Omega_t^d\cap\partial\Omega_t^n=\varnothing$, and
$\mathbf{n}$ is the outward unit normal on $\partial\Omega_t^n$. 

\section{The baseline solution for a growing strip}
We show the process of solving for the baseline symmetric solution of the full system in the simplest possible geometry, which is set up in Sec.~\ref{2A}. We then provide the details of the solution in Sec.~\ref{2B} and summarize the solution and the resulting baseline stress dynamics in Sec.~\ref{2C}. 

\subsection{The 2D geometry}
\label{2A}
We consider a growing strip with reference configuration
$\Omega_0=(-\infty,\infty)\times[0,1]$ in two-dimensional Cartesian
coordinates. The motion map is denoted by
\[
\mathbf{x}(\mathbf{X},t)=\left(x\left(X,Y,t\right),y\left(X,Y,t\right)\right)
\]
with the initial condition
\begin{equation}
\label{xyinitial}
x(X,Y,0)=X, \qquad  y(X,Y,0)=Y.
\end{equation}The bottom boundary $Y=0$ is clamped, while the top boundary $Y=1$ is free. Thus, the boundary conditions are
\begin{empheq}[left=\empheqlbrace]{align}
\label{xybc}
&x(X,0,t)=X,
\quad
y(X,0,t)=0,
\quad &\text{at } Y=0,
\\
\label{pkbc}
&\Pi_{12}(X,1,t)=0,
\quad
\Pi_{22}(X,1,t)=0,
\quad &\text{at } Y=1.
\end{empheq}
We assume the initial right Cauchy--Green growth tensor  \begin{equation}
\mathbf{C}_g(X,Y,0)=\mathbf{I}.\label{growthinitial}
\end{equation}
\subsection{Solution procedure}
\label{2B}
We denote the baseline solution of any state variable $\Box$ by $\Box^0$. By symmetry, all baseline variables are independent of $X$, except for the horizontal motion, for which
\begin{equation}x^0(X,Y,t)=X.\end{equation} Before solving Eqs.(\ref{thesystem1})-(\ref{thesystem3}), we observe that both the baseline deformation gradient \begin{equation}\mathbf{F}^0=\operatorname{diag}\left(1,\frac{\partial{y^0}}{\partial{Y}}\right)
\label{baseF}\end{equation}
and right Cauchy--Green growth tensor 
$\mathbf{C}^0_g = \operatorname{diag}\left(G_1,G_2\right)$
are diagonal. From Eqs.(\ref{implied_non-dimensional}) and (\ref{growthinitial}), we have $\det(\mathbf{C}^0_g)=G_1G_2 = \exp(2t)$, with time rescaled by the growth rate. Thus we write
\begin{equation}\mathbf{C}^0_g = \operatorname{diag}\left(G_1,\exp(2t)/G_1\right)\label{baseCg}\end{equation} with $G_1$ to be determined.

Using Eqs.(\ref{baseF}) and (\ref{baseCg}), the stress balance Eq.(\ref{thesystem1}) is reduced to 
\begin{equation}\label{1stpk}\frac{\partial \Pi^0_{22}}{\partial Y} = 0,\quad\text{ with } \quad \Pi_{11}^0 = e^{t}\left({G^{-1}_1} - {P^0}\right),  \quad\Pi_{12}^0 =\Pi_{21}^0 =0, \quad\Pi_{22}^0 = e^{t}\left(\frac{\partial{y^0}}{\partial{Y}}{G_1}{e^{-2t}} - {P^0}\left(\frac{\partial{y^0}}{\partial{Y}}\right)^{-1}\right).\end{equation}Using Eq.(\ref{baseF}), the incompressibility condition in Eq.(\ref{thesystem2}) becomes \begin{equation}\nabla\cdot\mathbf{v}^0= \frac{\partial^2{y^0}}{\partial{t}\partial{Y}}\left(\frac{\partial{y^0}}{\partial{Y}}\right)^{-1} =1,\quad\text{ with }\quad\nabla\mathbf{v}^0=\dot{\mathbf{F}}^0\left(\mathbf{F}^0\right)^{-1}=\operatorname{diag}\left(0,\frac{\partial^2{y^0}}{\partial{t}\partial{Y}}\left(\frac{\partial{y^0}}{\partial{Y}}\right)^{-1}\right).\end{equation} 
Using this together with the initial condition $\frac{\partial{y^0}}{{\partial{Y}}}(X,Y,0)=1$ derived from Eq.(\ref{xyinitial}), we have
 \begin{equation}\label{dydY}\frac{\partial{y^0}}{\partial{Y}}(X,Y,t)=e^{t}.\end{equation} Integrating it from the fixed boundary along the $Y$- direction, using Eq.(\ref{xybc}), gives \begin{equation}{y^0}(X,Y,t)=e^{t}Y.\end{equation}

Substituting Eqs.~(\ref{baseF}) and (\ref{baseCg}) into Eq.~(\ref{thesystem3}), together with the initial condition (\ref{growthinitial}), we have
\begin{equation}
\label{g1eq}
\dot{G}_1 = G_1+\frac{\beta}{2}\left(1-G_1^2\right) \quad\text{ with } G_1(X,Y,0)=1.
\end{equation}
Thus $G_1$ is spatially independent and satisfies a constant-coefficient Riccati equation in time when $\beta>0$. The solution is given by
 \begin{equation}G_1(t)=
\frac{
y_1-c\,y_2\exp\!\left(-\sqrt{1+\beta^2}\,t\right)
}{
1-c\exp\!\left(-\sqrt{1+\beta^2}\,t\right)
},
\quad \text{with } y_1=\beta^{-1}+\sqrt{1+\beta^{-2}}, \text{ }y_2=\beta^{-1}-\sqrt{1+\beta^{-2}}, \text{ and }
c=\frac{1-y_1}{1-y_2}.
\end{equation}
In this case, $G_1$ reaches the steady state $y_1$ as $t\to\infty$. See the main text for the effect of $\beta$ in plots. When $\beta=0$, $G_1(t)=e^{t}$ grows exponentially. We will mainly focus on the case of $\beta>0$. 

Given the closed-form solutions for $G_1(t)$ and $y^0$, we determine the
pressure from the traction-free boundary condition. From
Eq.~(\ref{1stpk}) and $\Pi_{22}^0(X,1,t)=0$, we have
$\Pi_{22}^0(X,Y,t)\equiv0$. Substituting
$\partial y^0/\partial Y=e^t$ then gives
\begin{equation}
P^0(t)=G_1(t).
\end{equation}
Consequently,
\begin{equation}
\Pi_{11}^0(t)
=
e^t\left(G_1^{-1}-G_1\right).
\end{equation}

\subsection{Solution summary}
\label{2C}
First, we summarize the baseline solution with minimal state variables:

\begin{equation}
\left\{
\begin{aligned}
&x^0(X,Y,t)=X,
\qquad
y^0(X,Y,t)=\exp(t)Y.
\\
&\mathbf{C}_{g}^0(X,Y,t)=\operatorname{diag}\left(G_1,\exp(2t)/G_1\right), \text{ with } G_1(t)=\frac{
y_1-c\,y_2\exp\!\left(-\sqrt{1+\beta^2}\,t\right)
}{
1-c\exp\!\left(-\sqrt{1+\beta^2}\,t\right)
},\\
&\quad  y_1=\beta^{-1}+\sqrt{1+\beta^{-2}}, \quad y_2=\beta^{-1}-\sqrt{1+\beta^{-2}},\quad \text{and}\quad c=\frac{1-y_1}{1-y_2}.
\\
&P^0(X,Y,t)=G_1(t).
\end{aligned}
\right.
\end{equation}
Then, the baseline first Piola-Kirchhoff stress is given by \begin{equation}
\Pi^0_{11}(X,Y,t)
=
e^t
\left(
G_1^{-1}
-
G_1
\right), \quad \text{and}\quad \Pi^0_{12}(X,Y,t)=\Pi^0_{21}(X,Y,t)=\Pi^0_{22}(X,Y,t)=0
\end{equation}
and the baseline Cauchy stress 
\begin{equation}
{\sigma}_{11}^0(X,Y,t)
=
\left(
G_1^{-1}-G_1
\right), \quad \text{and}\quad {\sigma}^0_{12}(X,Y,t)={\sigma}^0_{21}(X,Y,t)={\sigma}^0_{22}(X,Y,t)=0
\end{equation}
 from the Piola transformation $\boldsymbol\Pi^0 =\det{\mathbf{F}^0}\boldsymbol\sigma^0{\mathbf{F}^0}^{-T}$.

For $\beta>0$, since the horizontal growth component converges to the steady state
\begin{equation}
G_{1,\text{ss}}
= y_1=
\beta^{-1}+\sqrt{1+\beta^{-2}},
\end{equation}
the horizontal Cauchy stress converges to the steady state
\begin{equation}
\sigma_{11,\text{ss}}^{0}=(G^{-1}_{1,\text{ss}}-G_{1,\text{ss}})
=
-{2}{\beta}^{-1}.
\end{equation}
We note that the horizontal first Piola--Kirchhoff stress does not converge to a
steady value since 
\begin{equation}
\Pi_{11}^0(t)
=
e^{t}\sigma_{11}^0(t),
\end{equation}
whose magnitude grows exponentially, reflecting the continual increase of dimension along the $Y$-direction. 
\section{Linear stability analysis} 
We show the linearization of the system, Eqs.~(\ref{thesystem1})--(\ref{thesystem3}), about the baseline solution at a specific time in Sec.~\ref{3A}. We develop the analytical solution for the linearized system in Sec.~\ref{3B}. We discuss the linear system of boundary conditions that gives the dispersion relation in Sec.~\ref{3C}. While essential results are displayed in the main text, we show the stabilizing effect and the selection of small wavenumbers by surface tension in Sec.~\ref{3D}, and flow patterns at different levels of fluidity in Sec.~\ref{3E}.
\subsection{Linearization}
\label{3A}
At a given time $t_0\geq0$, we introduce the following perturbation about the symmetric solution
\begin{equation}
\label{perturbxy}
x(X,Y,t)\sim X+\epsilon x^1(X,Y,t;t_0), \quad y(X,Y,t)\sim e^{t}Y+\epsilon y^1(X,Y,t;t_0)
\end{equation}
\begin{equation}
\label{perturbgrowth}
C_{g,11}(X,Y,t)\sim G_1(t)+\epsilon G_{11}(X,Y,t;t_0), C_{g,12}(X,Y,t)\sim \epsilon G_{12}(X,Y,t;t_0), C_{g,22}(X,Y,t)\sim e^{2t}/G_1(t)+ \epsilon G_{22}(X,Y,t;t_0),
\end{equation}
\begin{equation}
P(X,Y,t)\sim G_1(t)+ \epsilon P^1(X,Y,t;t_0),
\end{equation}
 where $|\epsilon|\ll 1$.
 
Consequently, the perturbed deformation gradient components and the  growth volumetric ratio is given by 
\begin{equation}
\label{perturbF1}
{F}_{11}(X,Y,t)\sim 1+\epsilon \partial{x^1(X,Y,t;t_0)}/{\partial X},\quad {F}_{12}(X,Y,t)\sim \epsilon \partial{x^1(X,Y,t;t_0)}/{\partial Y},
\end{equation}
\begin{equation}
\label{perturbF2}
{F}_{21}(X,Y,t)\sim \epsilon \partial{y^1(X,Y,t;t_0)}/{\partial X},\quad {F}_{22}(X,Y,t)\sim e^{t}+\epsilon \partial{y^1(X,Y,t;t_0)}/{\partial Y},
\end{equation}
\begin{equation}
J_{g}(X,Y,t)\sim e^{t}+\frac{\epsilon}{2}\left(G_1(t)^{-1}e^tG_{11}(X,Y,t;t_0)+G_1(t)e^{-t}G_{22}(X,Y,t;t_0)\right).
\end{equation}
Substituting them into the stress balance Eq.(\ref{thesystem1}) at $t=t_0$, we obtain the linear equations
\begin{equation}
\label{stress1}
\left(G_1+G_1^{-1}\right)e^{t_0}
\frac{\partial^2 x^1}{\partial X^2}+{G_1}{e^{-t_0}}
\frac{\partial^2 x^1}{\partial Y^2}
+
G_1
\frac{\partial^2 y^1}{\partial X\,\partial Y}
-
e^{t_0}\frac{\partial P^1}{\partial X}=
\frac{1}{2}{\left(1+G_1^{-2}\right)}e^{t_0}
\frac{\partial G_{11}}{\partial X}+
{e^{-t_0}}
\frac{\partial G_{12}}{\partial Y}+
\frac{1}{2}{\left({G_1^2-1}\right)}e^{-t_0}
\frac{\partial G_{22}}{\partial X},
\end{equation}
\begin{equation}
\label{stress2}
G_1
\frac{\partial^2 x^1}{\partial X\,\partial Y}
+
G_1^{-1}e^{t_0}
\frac{\partial^2 y^1}{\partial X^2}
+
2G_1e^{-t_0}
\frac{\partial^2 y^1}{\partial Y^2}
-
\frac{\partial P^1}{\partial Y}=
\frac{\partial G_{12}}{\partial X}+G_1^2e^{-2t_0}
\frac{\partial G_{22}}{\partial Y}.
\end{equation}
where $G_1$ denotes $G_1(t_0)$ and each perturbation variable $\Box$ denotes $\Box(X,Y,t_0;t_0)$. We arrange the two equations so that the elastostatic terms are on the left-hand side and the growth-remodeling terms are on the right-hand side. Thus, perturbations on growth tensor introduce additional force terms in the stress balance.

From Eqs.~(\ref{deformrate}), (\ref{perturbF1}), and (\ref{perturbF2}), we have 
\begin{equation}
\label{gradvdiag}
(\nabla\mathbf{v})_{11}\sim \epsilon
{\partial^2 x^1(X,Y,t;t_0)}/{(\partial X\,\partial t)},\quad (\nabla\mathbf{v})_{22}\sim 1+\epsilon e^{-t}\left({\partial^2 y^1(X,Y,t;t_0)}/{(\partial Y\,\partial t)}-\partial y^1(X,Y,t;t_0)/{\partial Y}\right),
\end{equation}
\begin{equation}
\label{gradvoffdiag}
(\nabla\mathbf{v})_{12}\sim \epsilon
e^{-t}{\partial^2 x^1(X,Y,t;t_0)}/{(\partial Y\,\partial t)},\quad (\nabla\mathbf{v})_{21}\sim \epsilon\left({\partial^2 y^1(X,Y,t;t_0)}/{(\partial X\,\partial t)}-\partial y^1(X,Y,t;t_0)/{\partial X}\right).
\end{equation}
Substituting Eq.(\ref{gradvdiag}) into the incompressibility condition, Eq.(\ref{thesystem2}), at $t=t_0$, gives
\begin{equation}
\label{incomp}
 \frac{\partial^2 x^1}{\partial X\,\partial t}\bigg|_{t=t_0}+ e^{-t_0}
\left(
\frac{\partial^2 y^1}{\partial Y\,\partial t}\bigg|_{t=t_0}
-
\frac{\partial y^1}{\partial Y}
\right)
=0.
\end{equation}


Finally, substituting Eqs.(\ref{perturbgrowth}), (\ref{perturbF1}), and (\ref{perturbF2}) into the growth evolution Eq.(\ref{thesystem3}) gives three independent equations (due to the symmetry of $\mathbf{C}_g$):
\begin{equation}
\label{G11}
G_1^{-1}\frac{\partial G_{11}}{\partial t}\bigg|_{t=t_0}-\dot{G}_1G_1^{-2}G_{11}=\beta\left(-\frac{1}{2}G_1^{-2}G_{11}+\frac{1}{2}G_1^{2}e^{-2t_0}G_{22}+
{G_1^{-1}}\frac{\partial x^1}{\partial X}-{G_1}e^{-t_0}
\frac{\partial y^1}{\partial Y}\right).
\end{equation}
\begin{equation}
\label{G12}
e^{-2t_0}\left({G_1}\frac{\partial G_{12}}{\partial t}\bigg|_{t=t_0}
-
{\dot{G}_1}G_{12}\right)=\beta
\left(
-e^{-2t_0}{G_{12}}+
G_1e^{-2t_0}\frac{\partial x^1}{\partial Y}
+
{G_1}{e^{-t_0}}\frac{\partial y^1}{\partial X}
\right).
\end{equation}

\begin{equation}
\label{G22}
e^{-2t_0}\left(G_1\frac{\partial G_{22}}{\partial t}\bigg|_{t=t_0}-
\left(2G_1-\dot{G}_1\right)G_{22}\right)=-\beta\left(-\frac{1}{2}G_1^{-2}G_{11}+\frac{1}{2}G_1^{2}e^{-2t_0}G_{22}+
{G_1^{-1}}\frac{\partial x^1}{\partial X}-{G_1}e^{-t_0}
\frac{\partial y^1}{\partial Y}\right).\end{equation}We note that the right-hand side of Eqs.(\ref{G11}) and (\ref{G22}) are exactly opposite. This is expected because from Eq.(\ref{implied_non-dimensional}), the linear-order of $\tr(\dot{\mathbf{C}}_g\mathbf{C}_g^{-1})$ is strictly zero, corresponding to the sum of Eq.(\ref{G11}) and Eq.(\ref{G22}):
\begin{equation}
\label{G11pG22}
G_1^{-1}\frac{\partial G_{11}}{\partial t}\bigg|_{t=t_0}-\dot{G}_1G_1^{-2}G_{11}+e^{-2t_0}\left(G_1\frac{\partial G_{22}}{\partial t}\bigg|_{t=t_0}-
\left(2G_1-\dot{G}_1\right)G_{22}\right)=0\end{equation}

\subsection{Solution}
\label{3B}
In general, Eqs.~(\ref{stress1})--(\ref{stress2}), (\ref{incomp}), and (\ref{G11})--(\ref{G22}), or equivalently Eqs.~(\ref{stress1})--(\ref{stress2}), (\ref{incomp}), (\ref{G11})--(\ref{G12}), and (\ref{G11pG22}) form a coupled linear PDE system, and motivate the following ansatz:
 \begin{equation}
 \label{ansatzxy}
x^1(X,Y,t;t_0):=e^{q(t-t_0)}\cos(kX)u(Y;t_0), \quad y^1(X,Y,t;t_0):=e^{q(t-t_0)}\sin(kX)v(Y;t_0),
\end{equation}
 \begin{equation}
 \nonumber
G_{11}(X,Y,t;t_0):= e^{q(t-t_0)}\sin(kX)g_{11}(Y;t_0),  \quad G_{12}(X,Y,t;t_0):=e^{q(t-t_0)}\cos(kX)g_{12}(Y;t_0), \end{equation}
\begin{equation}
\label{ansatzG}
\quad\text{and}\quad G_{22}(X,Y,t;t_0):=e^{q(t-t_0)}\sin(kX)g_{22}(Y;t_0),
\end{equation}
 \begin{equation}
 \label{ansatzP}
P^1(X,Y,t;t_0):=e^{q(t-t_0)}\sin(kX)p(Y;t_0).
\end{equation} We are interested in solutions for which the growth rate of the perturbation $q$ is larger than the growth rate $\gamma=1$, henceforth we assume
\begin{equation}
\label{qlarger1}
q>1.
\end{equation}

We substitute them into Eqs.~(\ref{stress1})--(\ref{stress2}), (\ref{incomp}), (\ref{G11})--(\ref{G12}), and (\ref{G11pG22}), and obtain
\begin{equation}
\label{stress1an}
-k^2\left(G_1+G_1^{-1}\right)e^{t_0}u+{G_1}{e^{-t_0}}u''
+
kG_1
v'
-
ke^{t_0}p=
\frac{k}{2}{\left(1+G_1^{-2}\right)}e^{t_0}g_{11}+
{e^{-t_0}}
g'_{12}+
\frac{k}{2}{\left({G_1^2-1}\right)}e^{-t_0}g_{22},
\end{equation}
\begin{equation}
\label{stress2an}
-kG_1u'-k^2G_1^{-1}e^{t_0}v+2G_1e^{-t_0}v''-
p'=
-k g_{12}+G_1^2e^{-2t_0}g'_{22},
\end{equation}
\begin{equation}
\label{incompan}
 -qku+ (q-1)e^{-t_0}v'=0,
\end{equation}
%
\begin{equation}
\label{G11an}
\left((q-1)G_1^{-1}+\frac{\beta}{2}\right)g_{11}-\frac{\beta}{2}G_1^{2}e^{-2t_0}g_{22}=-\beta\left(
k{G_1^{-1}}u+{G_1}e^{-t_0}
v'\right),
\end{equation}
\begin{equation}
\label{G12an}
\left(\left(q-1\right)+\frac{\beta}{2}\left(G^{-1}_1+G_1\right)\right)e^{-2t_0}g_{12}=\beta
\left(e^{-2t_0}u'+k{e^{-t_0}}v
\right),
\end{equation}
and
\begin{equation}
\label{G11pG22an}
\left((q-1)-\frac{\beta}{2}\left(G^{-1}_1-G_1\right)\right)G_1^{-1}g_{11}+\left(\left(q-1\right) +\frac{\beta}{2}\left(G^{-1}_1-G_1\right)\right)G_1e^{-2t_0} g_{22}=0,\end{equation}
respectively, where we have used $\dot{G}_1 = G_1+{\beta}/{2}\left(1-G_1^2\right)$ from Eq.(\ref{g1eq}). The prime represents the differentiation with respect to $Y$.

One can observe that $u$, $g_{11}$, $g_{12}$, and $g_{22}$  can be solved in terms of $v$, $v'$, and $v''$ from Eqs.(\ref{incompan})-(\ref{G11pG22an}):
\begin{align}
\label{vsubu}
u
&=
(ke^{t_0})^{-1}(1-q^{-1})\,v',
\\[1ex]
g_{11}
&=
-\frac{
\beta e^{-{t_0}}
\left(1-q^{-1}+G_1^2\right)
\left((q-1)+\frac{\beta}{2}(G^{-1}_1-G_1)\right)
v'
}{
(q-1)\left((q-1)+\frac{\beta}{2}(G^{-1}_1+G_1)
\right)
},
\\[1ex]
\label{g_12}
g_{12}
&=
\frac{
\beta (ke^{{t_0}})
\left(
\,v
+
(ke^{{t_0}})^{-2}(1-q^{-1})v''
\right)
}{
\left((q-1)+\frac{\beta}{2}(G_1^{-1}+G_1)
\right)
},
\\[1ex]
\label{vsubg22}
g_{22}
&=
\frac{
\beta e^{{t_0}}
\left(1-q^{-1}+G_1^2\right)
\left((q-1)-\frac{\beta}{2}(G^{-1}_1-G_1)\right)
v'
}{
G_1^2(q-1)
\left((q-1)+\frac{\beta}{2}(G^{-1}_1+G_1)
\right)
}
\end{align}
which requires the two conditions
\begin{equation}
\nonumber
q-1\neq 0, \quad \text{and}\quad (q-1)+\frac{\beta}{2}(G^{-1}_1+G_1)\neq0,
\end{equation}
which are implied by $q>1$, $\beta>0$, and $G_1\geq 1$.


Differentiating Eq.~(\ref{stress1an}) and substituting the resulting expression for \(p'\), as well as Eqs.~(\ref{vsubu})-(\ref{vsubg22}) into Eq.~(\ref{stress2an})  yields the fourth-order ODE
\begin{equation}
v^{(4)}-\left(\frac{k^2 e^{2{t_0}}}{G^2_1}+\frac{
k^2 e^{2{t_0}}
\left((q-1)+\frac{\beta}{2}\left(G_1^{-1}-G_1\right)\right)
}{
{\left(1-{q}^{-1}\right)}\,
\left({(q-1)-\frac{\beta}{2}\left(G_1^{-1}-G_1\right)}\right)
}\right)v''+\left(\frac{k^2 e^{2{t_0}}}{G^2_1}\right)\left(\frac{
k^2 e^{2{t_0}}
\left((q-1)+\frac{\beta}{2}\left(G_1^{-1}-G_1\right)\right)
}{
{\left(1-{q}^{-1}\right)}\,
\left({(q-1)-\frac{\beta}{2}\left(G_1^{-1}-G_1\right)}\right)
}\right)v
=0
\end{equation}
with the general solution
\begin{equation}
\label{general}
v(Y)=c_1e^{-\lambda_1Y}+c_2e^{\lambda_1Y}
+c_3e^{-\lambda_2Y}+c_4e^{\lambda_2Y},
\end{equation}
with
\[
\lambda_1=\frac{k e^{t_0}}{G_1},
\qquad
\lambda_2=
\frac{
k e^{t_0}
\sqrt{(q-1)+\frac{\beta}{2}\left(G_1^{-1}-G_1\right)}
}{
\sqrt{1-{q}^{-1}}\,
\sqrt{(q-1)-\frac{\beta}{2}\left(G_1^{-1}-G_1\right)}
}
\]
when $\lambda_1\neq\lambda_2$ with the coefficients $\begin{pmatrix}
c_1 & c_2&  c_3 &c_4
\end{pmatrix}$ to be determined.  Given $q>1$ and $G_1\geq 1$,  the denomenator of $\lambda_2$ is always real. We find that the dispersion relation exists only when $\lambda_2$ is real which implies
\begin{equation}
(q-1)+\frac{\beta}{2}\left(G_1^{-1}-G_1\right)>0 \Longleftrightarrow\  q > 1-\frac{\beta}{2}\left(G_1^{-1}-G_1\right)
=1-\frac{\beta}{2}\sigma_{11}^0.
\end{equation}
We will also encounter marginal cases when $\lambda_1=\lambda_2$, and the general solution is instead given by 
\begin{equation}
\label{margingeneral}
v(Y)=c_1e^{-\lambda_1Y}+c_2e^{\lambda_1Y}
+c_3 Y e^{-\lambda_1Y}+c_4Y e^{\lambda_1Y},
\end{equation}
with the constraint 
\begin{equation}
\label{constraint}
\frac{1}{G_1} = \frac{
\sqrt{(q-1)+\frac{\beta}{2}\left(G_1^{-1}-G_1\right)}
}{
\sqrt{1-{q}^{-1}}\,
\sqrt{(q-1)-\frac{\beta}{2}\left(G_1^{-1}-G_1\right)}
}.
\end{equation}
This marginal case does not introduce any discontinuity of the dispersion relation.

\subsection{Boundary conditions and dispersion relation}
\label{3C}
The linearization of the fixed boundary conditions in Eq.~(\ref{xybc}) gives \begin{eqnarray}
\label{bcxy}
x^1\bigg|_{Y=0}=0, \quad \text{and} \quad y^1\bigg|_{Y=0}=0,
\end{eqnarray}
which, by using Eq.(\ref{ansatzxy}), leads to 
\begin{eqnarray}
\label{bcxymiddle}
v'\bigg|_{Y=0}=0, \quad \text{and} \quad v\bigg|_{Y=0}=0.
\end{eqnarray}

For the free-boundary conditions in Eq.(\ref{pkbc}), we have 
\begin{eqnarray}
\label{stressbc1}
\left(G_1 e^{-t_0}\frac{\partial x^1}{\partial Y}+G_1\frac{\partial y^1}{\partial X}\right)\bigg|_{Y=1}&=&G_{12}e^{-t_0}\bigg|_{Y=1}\\
\label{stressbc2}
\left(2G_1e^{-t_0}\dfrac{\partial y^1}{\partial Y}-P^1\right)\bigg|_{Y=1}&=&G^2_1 e^{-2t_0}G_{22}\bigg|_{Y=1},
\end{eqnarray}
which, using Eqs.(\ref{ansatzG}) and (\ref{ansatzP}), gives
\begin{eqnarray}
\label{stressbc1middle}
G_1\left( \left(ke^{t_0}\right)^{-1}u'+v\right)|_{Y=1}&=&g_{12}(ke^{t_0})^{-1}|_{Y=1}\\
\label{stressbc2middle}
\left(2G_1\left(ke^{t_0}\right)^{-1}v'-\left(k^{-1}p\right)\right)|_{Y=1}&=&G^2_1 \left(ke^{t_0}\right)^{-1} \left(e^{-t_0}g_{22}\right)|_{Y=1}.
\end{eqnarray}
By substituting Eq.(\ref{vsubu}) into the left-hand side of Eq.(\ref{stressbc1middle}), we obtain the expression $G_1\left( v+(ke^{t_0})^{-2}(1-q^{-1})v''\right)|_{Y=1}$, which shares the same cofactor $\left( v+(ke^{t_0})^{-2}(1-q^{-1})v''\right)$ with its right-hand side, as can be seen from Eq.(\ref{g_12}). Since other factors are nonzero and do not cancel each other in general, we simplify Eq. (\ref{stressbc1middle}) to 
\begin{equation}
\label{stressbc1new}
\left(\left(k e^{t_0} \right)^2 v
+(1-q^{-1})v''\right)|_{Y=1}=0.
\end{equation}

Further, we rewrite Eq.(\ref{stressbc2middle}) as
\begin{equation}
\label{stressbc2new}
\left.
\begin{aligned}
&\left(G_1+G_1^{-1}\right)u
+G_1\left(ke^{t_0}\right)^{-1}
\left(v'-\left(ke^{t_0}\right)^{-1}u''\right) \\
&\qquad
+\frac{1}{2}\left(1+G_1^{-2}\right)k^{-1}g_{11}
+\left(ke^{t_0}\right)^{-2}g'_{12}
-\frac{1}{2}\left(G_1^2+1\right)
\left(ke^{t_0}\right)^{-1}e^{-t_0}g_{22}
\end{aligned}
\right|_{Y=1}
=0
\end{equation}
by using 
\begin{equation}
\label{stressp}
k^{-1}p=-\left(G_1+G_1^{-1}\right)u+{G_1}\left(ke^{t_0}\right)^{-1}\left(\left(ke^{t_0}\right)^{-1}u''+v'\right)-\frac{1}{2}{\left(1+G_1^{-2}\right)}\left(k^{-1}g_{11}\right)-
\left(ke^{t_0}\right)^{-2}
g'_{12}-
\frac{1}{2}{\left({G_1^2-1}\right)}\left(ke^{t_0}\right)^{-1}\left(e^{-t_0}g_{22}\right)
\end{equation}
derived from Eq.(\ref{stress1an}).

Before constructing the dispersion relation, we note that the wavenumber $k$ and the baseline thickness ratio $e^{t_0}$ (current thickness of the baseline strip relative to its reference thickness) always appear through the combination $ke^{t_0}$ in the boundary conditions, Eqs.~(\ref{bcxymiddle}), (\ref{stressbc1new}), and (\ref{stressbc2new}), as well as in the general solution, Eq.~(\ref{general}). Imposing these boundary conditions therefore yields the linear system \begin{equation}
\label{clinear}
\mathbf{M}(ke^{t_0},q,\beta,G_1)
\begin{pmatrix}
c_1 & c_2 & c_3 & c_4
\end{pmatrix}^T
=\mathbf{0}.
\end{equation}
Furthermore, because $G_1$ and $\sigma_{11}^0=G_1^{-1}-G_1$ have a one-to-one correspondence for $G_1\geq1$, or equivalently $\sigma_{11}^0\leq0$, we can equivalently express the coefficient matrix as
$\tilde{\mathbf{M}}(ke^{t_0},q,\beta,\sigma_{11}^0)$.

A nontrivial perturbed solution exists only if
\begin{equation}
\label{dispersion}
\det\mathbf{M}(ke^{t_0},q,\beta,G_1)=0,
\quad\text{or equivalently}\quad
\det\tilde{\mathbf{M}}(ke^{t_0},q,\beta,\sigma_{11}^0)=0,
\end{equation}
which gives the dispersion relation for $q$. In the main text, we analyze the effects of fluidity $\beta$, the accumulated growth stretch $G_1$ (or stress $\sigma_{11}^0$), and the rescaled wavenumber $ke^{t_0}$ on $q$ by solving Eq.~(\ref{dispersion}) numerically in Mathematica.

\subsection{Dispersion relation with surface tension}
\label{3D}
We show stabilizing effect and wavenumber selection from a surface tension at the free boundary. More precisely, in 2D, we define a constant tension $\Lambda$ along the free edge $Y=1$, which modifies the boundary condition of $\Pi_{22}(X,1,t)$ in Eq.~(\ref{pkbc})
\begin{equation}
\label{pkbcc}
\Pi_{22}(X,1,t)=\Lambda \frac{d{\boldsymbol{\tau}}}{dX}\cdot\mathbf{N}
\quad \text{at } Y=1,
\end{equation}
where $\mathbf{N}=(0,1)$ and $\boldsymbol{\tau}$ is the unit tangent along the free-boundary curve parameterized by $X$. 
Here, the derivative is taken with respect to the reference coordinate $X$,
rather than the current arclength $s$, because $\boldsymbol{\Pi}\mathbf{N}$
is the nominal traction per unit reference length. Its relation to the
traction per unit current length
$\boldsymbol{\sigma}\mathbf{n}
=\Lambda\frac{d\boldsymbol{\tau}}{ds}$
is given by
\begin{equation}
\boldsymbol{\Pi}\mathbf{N}
=
\boldsymbol{\sigma}\mathbf{n}\frac{ds}{dS}
=
\Lambda\frac{d\boldsymbol{\tau}}{dS}
=
\Lambda\frac{d\boldsymbol{\tau}}{dX},
\end{equation}
where $S$ denotes the reference arclength and $dS=dX$ along the reference
free boundary.

The linearization of Eq.~(\ref{pkbcc}) modifies Eq.~(\ref{stressbc2}) to
\begin{eqnarray}
\left(2G_1e^{-t_0}\dfrac{\partial y^1}{\partial Y}-P^1\right)\bigg|_{Y=1}&=&\left(G^2_1 e^{-2t_0}G_{22}+\Lambda \dfrac{\partial^2 y^1}{\partial X^2}\right)\bigg|_{Y=1},
\end{eqnarray}
which further modifies Eq.~(\ref{stressbc2new}) to
\begin{equation}
\label{stressbc2newnew}
\left.
\begin{aligned}
&\left(G_1+G_1^{-1}\right)u
+G_1\left(ke^{t_0}\right)^{-1}
\left(v'-\left(ke^{t_0}\right)^{-1}u''\right)+\Lambda k v \\
&\qquad
+\frac{1}{2}\left(1+G_1^{-2}\right)k^{-1}g_{11}
+\left(ke^{t_0}\right)^{-2}g'_{12}
-\frac{1}{2}\left(G_1^2+1\right)
\left(ke^{t_0}\right)^{-1}e^{-t_0}g_{22}
\end{aligned}
\right|_{Y=1}
=0.
\end{equation}
The modified boundary condition in Eq.~(\ref{stressbc2newnew}) results in the updated linear system for the nontrivial-solution coefficients,
\begin{equation}
\mathbf{M}(k,e^{t_0},q,\beta,G_1,\Lambda)
\begin{pmatrix}
c_1 & c_2 & c_3 & c_4
\end{pmatrix}^T
=\mathbf{0}.
\end{equation}

\begin{figure}[h]
\centering
\includegraphics[width=\columnwidth]{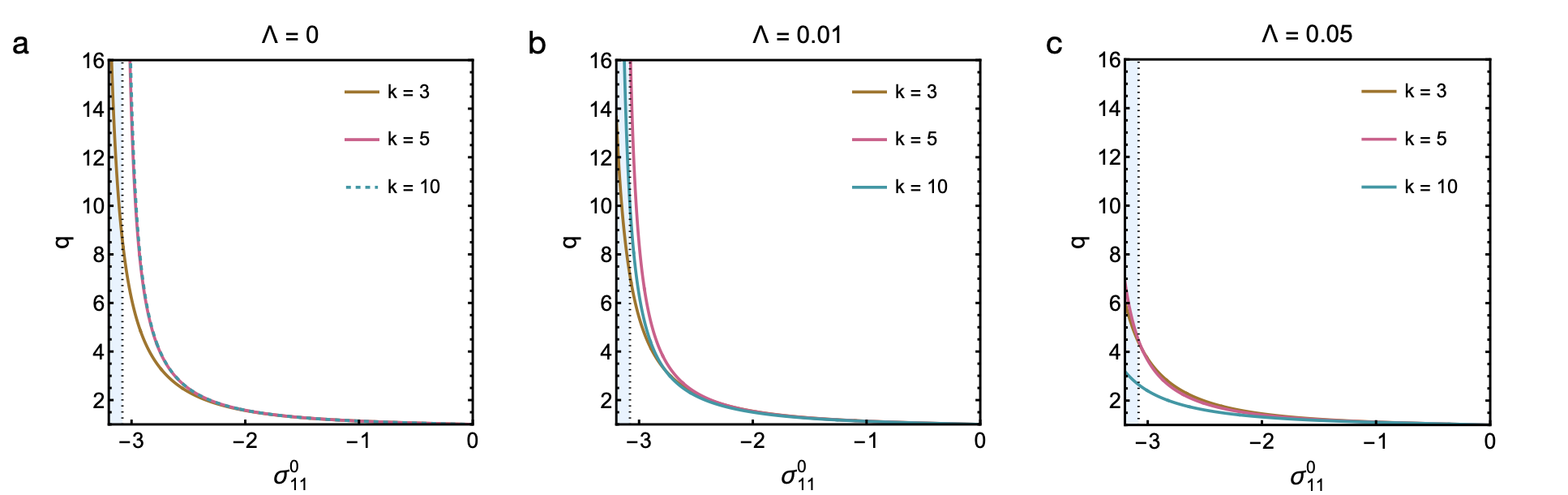}
\caption{\small The selection of $k$ with $\beta=0.2$ and $e^{t_0}=e\sim2.72$ as $\Lambda$ is varied. In (a), without tension ($\Lambda=0$), larger $k$ corresponds to larger $q$ at all values of $\sigma^0_{11}$. The overlap between $k=5$ and $k=10$ shows the saturation of the growth rate for $k\geq5$. Similarly, when we use $m=ke^t$ instead, we find saturation for $m\gtrsim12$. In (b,c), we show that increasing $\Lambda$ overall lowers the growth rate $q$, and the larger-$k$ curves drop faster than the lower-$k$ curves. The dotted line denotes the Biot threshold.}
\label{figS1}
\end{figure}

How $\Lambda$ affects the growth rate $q$ is demonstrated in Fig.S\ref{figS1} for $\beta=0.2$.

\subsection{Flow patterns under different levels of fluidity}
\label{3E}

For a finite instability growth rate $q$, the perturbation velocity remains finite during symmetry breaking, allowing the flow pattern and strain-rate field to be calculated using the coefficients $(c_1=1,c_2,c_3,c_4)$ determined from the linear system (\ref{clinear}). We visualize the perturbed boundary configuration with $\epsilon$ chosen such that the linear-order displacement remains small relative to the base flow. To visualize the flow pattern, we  plot the eigenvector field associated with the positive eigenvalue of the deviatoric strain-rate tensor $\mathbf{D}_D$, representing the maximal extension direction and rate, respectively. The strain-rate tensor $\mathbf{D}=\sym(\nabla\mathbf{v})$ is obtained using Eqs.~(\ref{gradvdiag}), (\ref{gradvoffdiag}), and (\ref{ansatzxy}), together with the relations between $u$ and $v$:
\begin{eqnarray}
\label{strainratediag}
D_{11}&=&(\nabla\mathbf{v})_{11}
\sim -\epsilon kq\,u(Y;t_0)\sin(kX)
=-\epsilon e^{-t_0}(q-1)v'(Y;t_0)\sin(kX),\\
D_{22}&=&(\nabla\mathbf{v})_{22}
\sim 1+\epsilon e^{-t_0}(q-1)v'(Y;t_0)\sin(kX),\\
D_{12}&=&D_{21}
=\frac{1}{2}\left[\left(\nabla\mathbf{v}\right)_{12}
+\left(\nabla\mathbf{v}\right)_{21}\right]\nonumber\\
&\sim&
\frac{\epsilon}{2}
\left[
q e^{-t_0}u'(Y;t_0)
+k(q-1)v(Y;t_0)
\right]\cos(kX)\nonumber\\
&=&\frac{\epsilon k(q-1)}{2}
\left[(ke^{t_0})^{-2}v''(Y;t_0)+v(Y;t_0)
\right]\cos(kX).
\label{strainrateoffdiag}
\end{eqnarray}

\begin{figure}[h]
\centering
\includegraphics[width=\columnwidth]{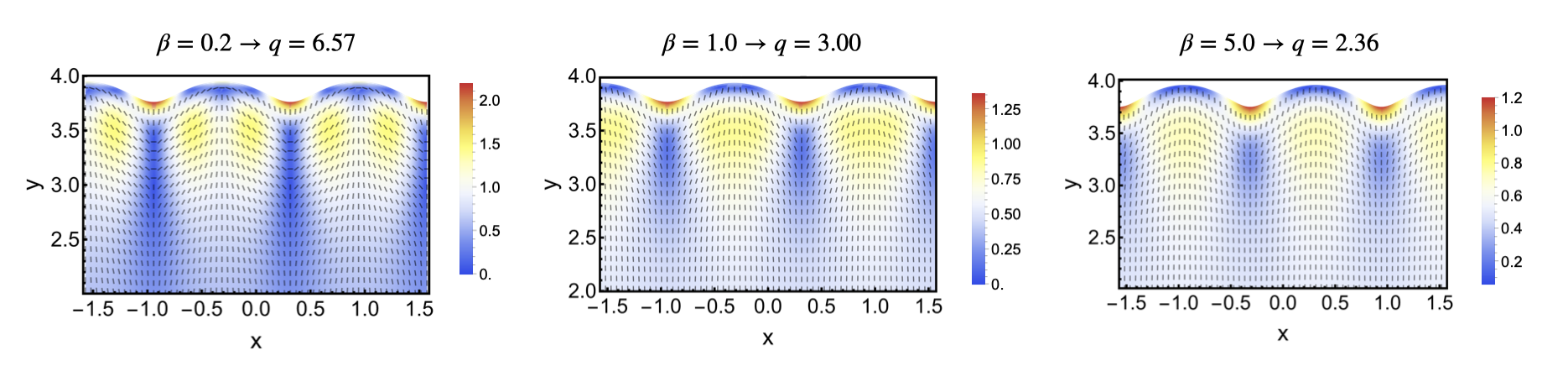}
\caption{\small The heat maps show the maximal extension rate ($\lambda^+$ of $\mathbf{D}_D$), while the short bars indicate the corresponding maximal-extension direction. At the same time $t=1.35$, increasing $\beta$ leads to a smaller $q$ and weaker localization of the indentations, as reflected by the less pronounced ridges in the extension-rate field.}\label{figS2}
\end{figure}

We show in Fig.S\ref{figS2} the flow patterns at $t=1.35$ for different values of $\beta$, corresponding to different values of $q$. Higher $q$ produces stronger localization of the periodic indentations.

\bibliography{refs}